\documentclass[a4paper,11pt]{article}
\pdfoutput=1 

\usepackage{jcappub} 

\usepackage[T1]{fontenc} 

\usepackage[dvipsnames]{xcolor}

\usepackage{fontawesome5}

\usepackage[section]{placeins}

\newcommand\pasa{\ref@jnl{PASA}}

\usepackage{enumerate}
\usepackage{hyperref}
\usepackage{url}
\usepackage{subcaption}
\usepackage[normalem]{ulem}
\title{Searching for dark-matter waves with PPTA and QUIJOTE pulsar polarimetry}

\author[a,b]{Andr\'es Castillo}
\author[a,b]{Jorge Martin-Camalich}
\author[a,b]{Jorge Terol-Calvo}
\author[c,d]{Diego Blas}
\author[e,f]{Andrea Caputo}
\author[a,b]{Ricardo Tanaus\'u G\'enova Santos}
\author[g]{Laura Sberna}
\author[a,b]{Michael Peel}
\author[a,b]{Jose Alberto Rubi\~no-Mart\'in}

\affiliation[a]{Instituto de Astrof\'isica de Canarias, C/ V\'ia L\'actea, s/n
E38205 - La Laguna, Tenerife, Spain}
\affiliation[b]{Universidad de La Laguna, Dpto. Astrof\'isica, E38206 - La Laguna, Tenerife, Spain}
\affiliation[c]{Grup de F\'isica Te\`orica, 
Departament  de  F\'isica, Universitat  Aut\`onoma  de  Barcelona,   Bellaterra, 08193 Barcelona, Spain}
\affiliation[d]{Institut de Fisica d’Altes Energies (IFAE), The Barcelona Institute of Science and Technology, Campus UAB, 08193 Bellaterra  (Barcelona), Spain}
\affiliation[e]{School of Physics and Astronomy, Tel-Aviv University, Tel-Aviv 69978, Israel}
\affiliation[f]{Department of Particle Physics and Astrophysics, Weizmann Institute of Science, Rehovot 7610001, Israel}
\affiliation[g]{Max Planck Institute for Gravitational Physics (Albert Einstein Institute), Am Mu\"{u}hlenberg 1, 14476 Potsdam, Germany}

\emailAdd{acastillo@iac.es}
\emailAdd{jcamalich@iac.es}
\emailAdd{jorgetc@iac.es}

\abstract{The polarization of photons emitted by astrophysical sources might be altered as they travel through a
dark matter medium composed of ultra light axion-like particles (ALPs). In particular, the coherent oscillations of the ALP background in the galactic halo induce a periodic change on the 
polarization of the electromagnetic radiation emitted by local sources such as pulsars. Building up on previous works, we develop a new, more robust, analysis based on the generalised Lomb-Scargle periodogram to search for this periodic signal in the emission of the Crab supernova remnant observed by the QUIJOTE MFI instrument and 20 Galactic pulsars from the Parkes Pulsar Timing Array (PPTA) project. We also carefully take into account the stochastic nature of the axion field, an effect often overlooked in previous works.
This refined analysis leads to the strongest limits on the axion-photon coupling 
for a wide range of dark matter masses spanning $10^{-23}\text{ eV}\lesssim m_a\lesssim10^{-19} \text{ eV}$. Finally, we survey possible optimal targets and the potential sensitivity to axionic dark-matter in this mass range that could be achieved using pulsar polarimetry in the future.

}

\begin{document}
\maketitle

\section{Introduction}

The fundamental nature of dark matter (DM) remains one of the major mysteries of modern physics~\cite{Bertone:2010zza}.
One of the most promising candidates of particle DM is the axion~\cite{Preskill1983,Abbott:1982af,Dine:1982ah,Marsh_2016}, whose first incarnations were predicted as a consequence of the Peccei-Quinn solution to the strong \emph{CP}-problem~\cite{Peccei:1977hh,Wilczek:1977pj,Weinberg:1977ma}. Equally interesting is the larger framework of axion-like particles (ALPs),  pseudoscalar particles with Standard Model (SM) couplings resembling  those of the QCD axion~\footnote{Here we employ usual terminology in high-energy physics, where QCD refers to Quantum Chromo-Dynamics (the fundamental theory of the strong interactions) and $CP$ stands for the combined charge-conjugation ($C$) and parity ($P$) symmetry transformation.}, but not necessarily related to the strong \emph{CP}-problem. As a consequence,  their mass and decaying constant can be decoupled, widely broadening the possible phenomenological consequences.   

A characteristic feature of ALPs is that they can be extremely light ($\sim$ 10$^{-22}$ eV), becoming the prototypical example of bosonic Ultra Light Dark Matter (ULDM), also called fuzzy dark matter or wave dark matter~\cite{Peebles:2000yy,Goodman:2000tg,Hu:2000ke,Hui:2016ltb,Niemeyer:2019aqm,Hui:2021tkt}. Remarkably, this type of ultra light particles is naturally predicted in some extensions of the SM, including string theory~\cite{Svrcek:2006yi,Arvanitaki:2009fg}. These dark matter candidates are characterized by a large de Broglie wavelength, which suppresses structure at small  
scales~\cite{Hu:2000ke,Hui:2021tkt}, as guaranteed by the uncertainty principle. This property has been used as further motivation for these models, as this suppression may
address some difficulties of the cold dark matter paradigm at galactic scales~\cite{Weinberg:2013aya}. A prominent example is
the core-cusp problem, arising  from the discrepancy between the inferred dark matter density profiles of low-mass galaxies and the density profiles predicted by cosmological \texttt{N}-body simulations \cite{Deng_2018}. Other related observables where ULDM may play a critical role include: the Lyman-$\alpha$ forest  \cite{Vid_2017,Zhang_2018}, the abundance of high-redshift objects \cite{Menci_2017}, galactic rotation curves \cite{Bar:2018acw,Bar:2021kti} and the formation features  of the Milky Way disk \cite{Banik_2017}.

Another remarkable property of these DM candidates is that they can be easily (even ``naturally'') produced at the observed cosmological levels via the misalignment mechanism~\cite{Preskill1983, Turner:1983he, Abbott:1982af}. This relies on assuming that the field strength has a ``natural'' amplitude in the early universe (provided by, e.g., a phase transition)  and that after some time (depending on the  mass of the DM particle) they start to oscillate and behave as a cold dark matter fluid.
 
In this work we will focus on the ULDM parameter space of ALPs. One of the most interesting handles in their phenomenology is the  non-renormalizable interaction of ALPs with electromagnetic (EM) fields. This interaction might be studied in several kinds of experiments such as CAST~\cite{Cast2017}, or astrophysical phenomena like the supernova SN~1987A~\cite{Payez:2014xsa}. It also
  affects the emission, propagation and detection of EM waves from astrophysical objects (see e.g. Refs.~\cite{Raffelt:1996wa,Irastorza:2018dyq,pdg} for reviews). In this work, we will describe how the polarization properties of light are modified in its propagation through an external ALP (pseudoscalar) field. 
In particular, we focus on the modulation of the light polarization angle induced by ALPs in virial equilibrium in galactic structures~\cite{Carroll:1989vb,Harari:1992ea,Hui:2021tkt}. 
This effect has been searched for in, for instance, pulsars~\cite{Caputo:2019tms}, parsec-scale jets in active
galaxies~\cite{Ivanov:2018byi}, protoplanetary disks ~\cite{PDPAxion} or in observations of the CMB~\cite{Fedderke:2019ajk, Sigl:2018fba}. In the present manuscript we improve and refine the analysis of Ref.~\cite{Caputo:2019tms} using pulsars as light sources. Our analysis is novel in several ways:
\begin{itemize}
    \item First, we use generalized Lomb-Scargle periodograms. These are standard and powerful techniques used in astronomy to search for periodic signals in non-homogeneously distributed time series. This allows us to perform a robust analysis of the time dependence of the pulsar's polarization measurements constraining ALP-birefringence effects. (Similar techniques were applied to active galaxy sources in this context~\cite{Ivanov:2018byi}).
    \item Second, we use a more accurate model of the ULDM configuration. This takes into account the intrinsic stochastic nature of the amplitude of the field, which is relevant for the mass range of interest and experimental setup. 
    This is tackled in our work by a global analysis of the polarization measurements of 20 pulsars observed for $\sim4.5$ years by the Parkes Pulsar Time Array (PPTA)~\cite{Yan:2011bq}. Furthermore, we include data from the polarization studies of the Crab Pulsar (and supernova remnant) that span also $\sim 4.5$ years, performed by the QUIJOTE MFI instrument~\cite{RubinoMartin_2012,Genova-Santos:2015uia} to calibrate their measurements.
    \item Finally, we investigate the prospects to improve the current sensitivity of this search strategy. In particular, we indicate the optimal features of potential observational targets, which might allow us to start probing a larger parameter space of ALP DM in the near future. Still, our current analysis presents the strongest constraints that have been obtained so far for the range $10^{-23}\text{ eV}\lesssim m_a\lesssim10^{-19} \text{ eV}$ of ultralight ALP masses.  
    
\end{itemize}

Our work is organized as follows: In sec.~\ref{sec:ALPs}, we introduce the concept of ALP, its relevance as a cold dark matter candidate, and the birefringence effect. In the last part of the same section, we discuss the impact of the stochastic nature of the ALP oscillations, deferring a detailed quantitative analysis of this to Appendix~\ref{sec:AppVs}. In sec.~\ref{sec:data} we describe the datasets while in sec.~\ref{sec:analysis} we explain the techniques--based on periodograms--to search for ALP effects on the \emph{polarization} of radio EM waves emitted in pulsars. In the final part of sec.~\ref{sec:analysis}, the individual and combined searches using PPTA and QUIJOTE are presented. Finally, in sec.~\ref{sec:forecasts} we summarize some prospects of our studies.

\section{ALPs as Ultra Light Dark Matter and the birefringence of light}
\label{sec:ALPs}
 
\subsection{ALPs as cold dark matter}
\label{sec:ALPDM}

As mentioned above, one of the most appealing features of ALP dark matter is that it can easily be produced at the right abundance with the misalignment mechanism. The equation of motion of the ALP field $a$ in an isotropic and homogeneous expanding universe reads
\begin{equation}
    \ddot{a} + 3H\dot{a} + m_a^2 a = 0,
\end{equation}
where $m_{a}$ is the ALP mass, $H =(\rho_{\rm tot}/3 M_{\rm Pl}^2)^{1/2}$ is the Hubble parameter with $\rho_{\rm tot}$ being the total energy density in the Universe. At early times, the ALP field
is frozen. Then, when the Hubble parameter becomes comparable to $m_a$, it starts to oscillate and soon after the ALP energy density
redshifts as cold matter. At the onset of oscillations, which is characterized by the temperature $T_{\rm osc}$, the ALP number density is given by
\begin{equation}
    n_a(T_{\rm osc}) = \frac{1}{2}m_a  f_a^2 \theta_0^2,
\end{equation}
where $f_a$ is the axion decay constant and $\theta_0$ the initial misalignment angle, defined by the initial value of the axion field $a_0 = f_a \theta_0$. In the conventional cosmological scenario, the ALP starts to oscillate during radiation domination, and the evolution below $T_{\rm osc}$ is assumed to be adiabatic. In this case one can write the axion density parameter as~\cite{Arias:2012az, Blinov:2019rhb, Dror:2020zru,Hui:2016ltb} 
\begin{equation}
\label{eq:ALPDMabundance}
    \Omega_a h^2 \simeq 0.12 \Big(\frac{f_a \theta_0}{6 \times 10^{16} \rm GeV}\Big)^2\Big(\frac{m_a}{2\times 10^{-21} \rm eV}\Big)^{1/2}.
\end{equation}
The scaling above is easy to understand. A larger decay constant means a larger axion energy density at the start of the oscillations. Furthermore, the larger the mass, the earlier the axion starts redshifting as matter (i.e., slower than radiation), allowing for $\rho_a/\rho_{\rm tot}$ to grow over a longer period.

\subsection{ALP-induced birefringence}
\label{sec:ALPbirefringence}

\begin{figure}[t!]
\begin{center}
  \includegraphics[width=10cm]{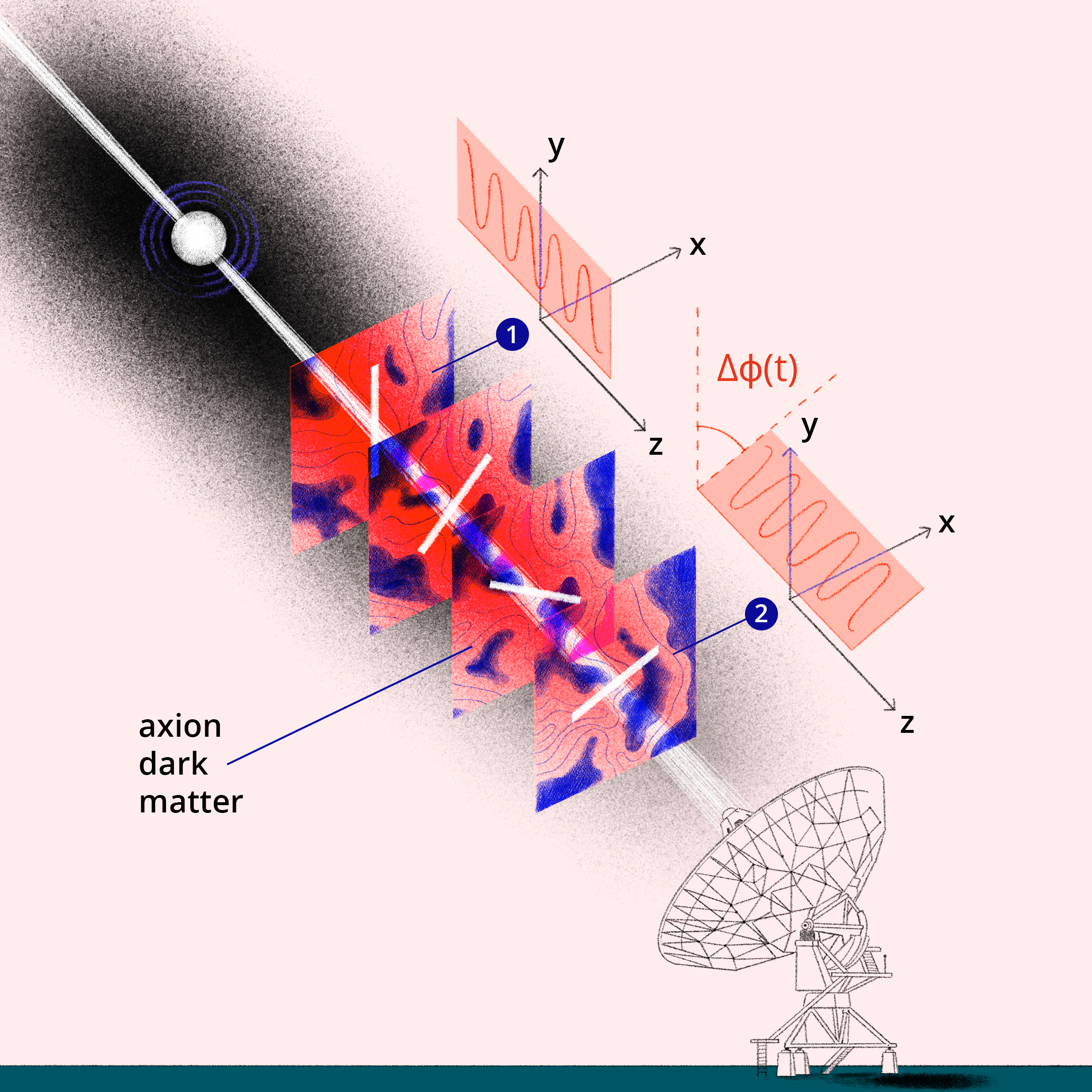}
\caption{Time-dependent birefringence induced by the axion DM field. Light is emitted by a given source (point 1) with a given polarization and received (point 2) with a different polarization $\Delta \phi(t)$. The effect is ``conservative'' in the sense that it depends exclusively on the difference between the values of the axion field at points 1 and 2. The time dependence is given by the coherent variation of the axion field, with frequency $\nu_c$. (Artist: \`Eve Barlier.)
\label{Fig:Diag}}
\end{center}
\end{figure}

The presence of an ALP relic  can lead to important phenomenological consequences. For example, its presence will modify the propagation of EM waves. In particular, the ALP medium would rotate the polarization plane of linearly polarized EM waves, such as those emitted by pulsars \cite{Carroll:1989vb,Harari:1992ea,Caputo:2019tms,Liu:2019brz}.
The origin of this phenomenon can be understood starting from the Lagrangian for an ALP field $a$, including its most relevant interactions to light. The latter has the following form

\begin{equation}
    \mathcal{L}=\frac{1}{4}F_{\mu\nu}F^{\mu\nu}+\frac{g_{a\gamma}}{4} aF_{\mu\nu}\tilde{F}^{\mu\nu}+\frac{1}{2}\left(\partial_{\mu}a\partial^{\mu}a-m_{a}^{2} a^2\right)\,, \label{eq:Lagrangian}
\end{equation}
where 
$F_{\mu\nu}$ is the EM stress tensor and $\Tilde{F}^{\mu\nu}=\varepsilon^{\mu\nu\rho\sigma}F_{\rho\sigma}/2$ is its dual. As follows from dimensional analysis, the coupling constant $g_{a \gamma}$ has units of inverse of mass. 
The field equation for the ALP field reads
\begin{equation}\label{eq:eoma}
\big(\Box{}+m_{a}^2\big)a+\frac{g_{a \gamma}}{4}a F_{\mu\nu} \tilde{F}^{\mu\nu}=0\,, 
\end{equation}
and goes together with the modified Maxwell's equations 
coupled to the ALP field 
\begin{align}
\partial_{\mu}F^{\mu\nu}+g_{a \gamma}\partial_{\mu}\left(\tilde{F}^{\mu\nu}a\right) &=0, \\
\partial_{\mu}\tilde{F}^{\mu\nu}&=0\,. 
\end{align}
The changes produced by the $g_{a \gamma}a$ term can be considered adiabatic at scales of the order of the EM wavelength, since we will work only in the limit $\partial_t a\ll \omega$ and $\vert\nabla a\vert\ll k$, $\omega$ being the frequency of the radio wave and $\mathbf{k}$  its momentum (wave-vector).  Then, as far as the propagation of the EM wave is concerned, $\nabla a$ and $\partial_{t}a$ are approximately constant.  Terms with first spatial and temporal derivatives squared scale with $\vert\nabla a\vert/k$ and $\partial_t a/ \omega$, respectively, so they are also small and neglected. This approach coincides with geometrical optics, where the EM wavelength does not change appreciably over the characteristic length scale of spatial variations of $a$ field. In this context, the EM field admits plane wave solutions with  angular frequencies for left and right circularly polarized waves, 
\begin{align}
\label{eq:ALPinducedPS}
    \omega_{\pm}\simeq k \pm \frac{1}{2}g_{a \gamma}\left(\partial_{t}a+\nabla a \cdot\widehat{\mathbf{k}}\right),
\end{align}
to lowest order in $g_{a\gamma}$~\cite{Carroll:1989vb,Harari:1992ea}. Hence, the ALP field acts as a birefringent medium due to the parity-violating nature of the light-axion interaction: a linearly polarized EM wave changes its polarization plane due to the phase shift induced between left- and right-handed circular polarization components in Eq.~\eqref{eq:ALPinducedPS}. When integrated along the trajectory of propagation of the EM wave, the previous effect yields a modification of the polarization angle by,
\begin{align}
\label{eq:polALPgen}
    \Delta \phi =\frac{g_{a\gamma}}{2} \int_{t_{s}}^{t_{o}} \frac{da}{dt} dt= \frac{g_{a\gamma}}{2}\Delta a.
\end{align}
Two important conclusions follow from this equation. First, this birefringent effect is independent of the frequency of the EM wave. Secondly, it is ``conservative'' in the sense that the net rotation of the polarization angle depends only on the difference of the values of the ALP field at the source and observer (labeled by times $t_{s}$ and $t_{o}$ respectively) and not on the values of the ALP field crossed by the EM wave in its trajectory\footnote{Note also that the light-axion interaction we are studying does not lead to modifications of the trajectories of the two helicities at the precision we are studying \cite{Blas:2019qqp,McDonald:2019wou}.}. We show in Fig.~\ref{Fig:Diag} a graphical representation of this phenomenon.

We now  describe the changes of the polarization angle of light caused by the difference in the  ALP field in the Solar system as compared to the local environment of  pulsars as a function of time~\cite{Ivanov:2018byi,Caputo:2019tms,Liu:2019brz}. In both cases, the dark matter distribution can be considered as a  collection of free waves \cite{Hui:2016ltb,Foster:2017hbq,Foster:2020fln},
\begin{align}
      a(\vec x,t)=\int d^3 k\, a(\vec{k}) e^{i(\omega t-\vec{k}\cdot \vec{x})}+c.c. \, , \label{eq:distrib}
\end{align}
where $a(\vec k)$ incorporates the virialized properties of the distribution (in particular its coldness) and includes random phases for each wavenumber.
From the dispersion relation derived from \eqref{eq:eoma} at first order in the background fields, and from the fact that the virialized distribution is non-relativistic (cold), one can write this field configuration at \emph{any} time $t$ and position $\vec x$ as,
 \begin{align}
    a(\vec x,t) = a_0(\vec x)\cos\left(m_a t + \delta(\vec x)\right), 
 \end{align}
where $a_{0}(\vec x)$ and $\delta(\vec x)$ are the amplitude and phase of the field, respectively, at the location $\vec x$. 
This description is valid for times smaller than the coherence time
\begin{equation}
\label{eq:coherencet}
\tau_c=(m_a \sigma^2)^{-1}\simeq2\times10^5\left(\frac{m_a}{10^{-22}~\text{eV}}\right)^{-1}\left(\frac{\sigma}{10^{-3}}\right)^{-2}~\text{yr},
\end{equation}
where $\sigma\sim 10^{-3}$
is the dispersion velocity of dark matter in our Galaxy~\cite{Green:2017odb} and for distances smaller than the coherence length
 \begin{equation}
 \label{eq:coherenced}
     l_{c} =\left(m_{a}\sigma\right)^{-1}\simeq 65  \left(\frac{m_{a}}{10^{-22} \text{ eV}}\right)^{-1}\left(\frac{\sigma}{10^{-3}}\right)^{-1} {\rm pc},
 \end{equation}
 around the position $\vec x$. Both properties hinge on the virialized axion field~\eqref{eq:distrib} containing modes with different nonrelativistic velocities and random phases, which interfere and decohere the net field. 
To summarize, within the previous limits, the distribution behaves as a coherent field oscillating at  the nominal Compton frequency
\begin{equation}
\label{eq:ComptonFreq}
\nu=\nu_c=\frac{m_a}{2\pi}.
\end{equation}
Finally, one can relate the axionic field to the local dark matter energy density as derived from the stress energy tensor of an oscillating scalar field~\cite{Ivanov:2018byi}, 
\begin{equation}
\label{eq:rho_a}
    \rho_{\rm DM}=\frac{1}{2}m_a^2 \langle a^2\rangle\,,
\end{equation}
where the average corresponds to time averages at cosmological time scales, which are much longer than $\tau_c$ in \eqref{eq:coherencet} for the masses of interest. 

In the case of interest in our work, the relevant time scale is set by the duration of the observations, which is $\ll \tau_c$. In this situation, the connection between the axionic field amplitude and $\rho_{\rm DM}$ in \eqref{eq:rho_a} acquires a stochastic nature, with the amplitude of the axion field $a_{0,i}$ following a Rayleigh distribution in a given coherence domain (labeled by $i$)~\cite{Foster:2017hbq,Centers:2019dyn}. For the sake of clarity, we write $a_{0,i}=\sqrt{2 \rho_i}m_a^{-1}\alpha_i$, where we factored out the dependence on the local density of dark matter $\rho_i$ and the ALP mass, and $\alpha_i$ is a random variable following the probability distribution function (PDF)~\cite{Foster:2017hbq}\footnote{Alternatively, one can directly sample the energy density using the distribution~\cite{Knirck:2018knd} $p(\rho) = \frac{1}{\bar{\rho}}\exp(-\rho/\bar{\rho})$, where $\bar{\rho}$ indicates the average energy density (locally we have $\bar{\rho}\simeq 0.4\, \rm GeV/cm^3$).} 
\begin{equation}
\label{eq:Rayleigh}
p(\alpha)=\alpha \exp\left(-\frac{\alpha^2}{2}\right).
\end{equation}

With the previous ingredients, we can finally study the effect of the axion dark-matter field on the polarization of the light received from a given pulsar as a function of time, taking into account the stochastic nature of the field. From Eq.~\eqref{eq:polALPgen}, 
one finds
\begin{equation}
\label{eq:polALP1}
\Delta\phi(t)=\frac{g_{a\gamma}}{\sqrt{2}m_a}\left[\sqrt{\rho_o}\alpha_o\cos(m_a t+\delta_o)-\sqrt{\rho_s}\alpha_s\cos(m_a (t-T)+\delta_s)\right],
\end{equation}
where $t$ is the local time measured since the beginning of the observations (which defines $t=0$), $T$ is the time light takes to travel from the pulsar to Earth, and  $\delta_i$, $\rho_i$ and $\alpha_i$ are the phases, local dark-matter density and stochastic amplitude variable, respectively, at a given source ($i=s$) or observer ($i=o$). The variables $\delta_i$ are random,  with a flat PDF in $[0,2\pi]$. 
The expression~\eqref{eq:polALP1} can be compactly reorganized as
\begin{equation}
\label{eq:polALP2}
    \Delta\phi(t)=\phi_a\cos(m_a t+\varphi_a),
\end{equation}
where 
\begin{equation}
\label{eq:phi0}
\phi_a=\frac{g_{a\gamma}}{\sqrt{2}m_a}\left(\rho_o\alpha_o^2+\rho_s\alpha_s^2-2\sqrt{\rho_o\rho_s}\alpha_o\alpha_s\cos\Delta\right)^{1/2},
\end{equation} 
with $\Delta=m_a T+\delta_o-\delta_s$. In terms of the typical values for the relevant quantities $(\alpha_s=\alpha_o=1,\, \Delta=\pi,\, \rho_s=\rho_o=\rho_{\rm DM})$ we get

\begin{align}
  \phi_{a} =  4.48^{\circ}\, \left(\frac{g_{a\gamma}}{10^{-12}\text{ GeV}^{-1}}\right)\left(\frac{m_{a}}{10^{-22} \text{ eV}}\right)^{-1}\left(\frac{\rho_{DM}}{1 \text{ GeV cm}^{-3}}\right)^{1/2}.
\end{align}

The expression of the phase $\varphi_a$ in Eq.~\eqref{eq:polALP2} is not relevant, since it can eventually be considered as a random phase with a flat distribution in $[0,2\pi]$ \textit{and} uncorrelated with $\phi_a$\footnote{To be precise, $\varphi_a$ is a function of the random variables $\delta_i$ and $\alpha_i$ (and of $\rho_i$), but it can be treated as uniformly distributed. We have checked this claim numerically for different instances of the parameters in \eqref{eq:polALP1}.}. The argument $\Delta$  in Eq.~\eqref{eq:phi0} can also be treated as a random phase even if the pulsar and Earth belong to the same coherence domain (i.e. when $\delta_s=\delta_o$). In this case, the phase only depends on $m_a T$, which should also be treated as a random variable since the uncertainty in the measurement of the distance  typically spans several cycles of the ALP field.

\begin{figure}[t!]
\begin{center}
  \includegraphics[width=15cm]{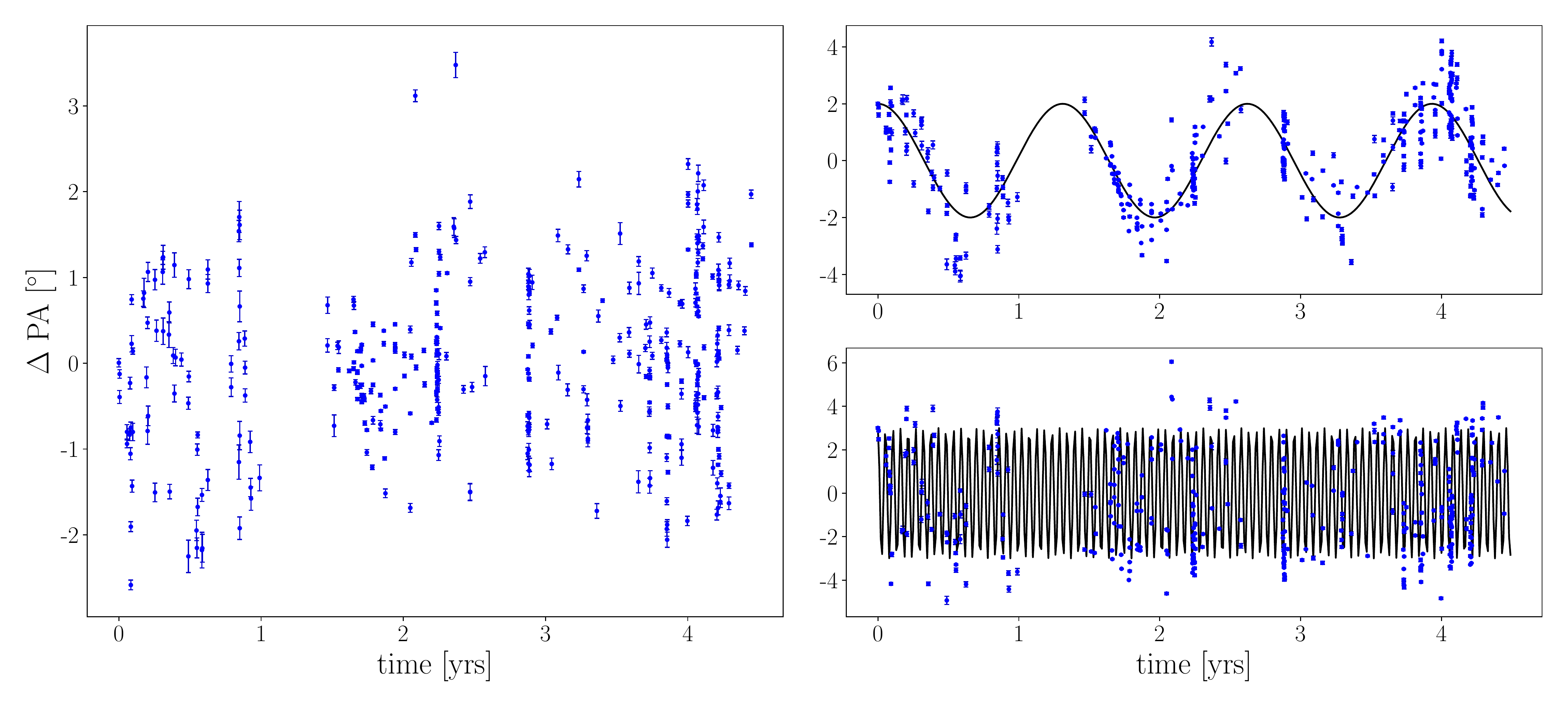}
\caption{On the left, the time series of the pulsar \texttt{J0437-4715} as measured. On the right, the same time series with the addition of an ALP signal, in blue. The theoretical signals are shown in black. The top panel shows a signal with ALP mass $m_a=10^{-22}$ eV, corresponding to a frequency $\nu=2.16\times 10^{-3}$ days$^{-1}$, and an amplitude of 2 deg. The bottom one shows a signal with $m_a=5\times10^{-20}$ eV, corresponding to a frequency $\nu=1.07$ days$^{-1}$, and an amplitude of 3 deg.
\label{Fig:TimeSeries}}
\end{center}
\end{figure}

In summary, one can indirectly search for the interaction of light with the dark-matter waves of the ALP field by looking for  periodic variations\footnote{Note that, when looking for a periodic signal, one can discard any constant time shift. In particular, one can for example neglect the frequency dependent delay due to the interstellar medium dispersion. Nonetheless, we checked that this latter effect amounts to a negligible delay (less than a second) for all the sources we consider.} in $\Delta \phi(t)$, Eq.~\eqref{eq:polALP2}. In Fig. \ref{Fig:TimeSeries} we can see how different ALP signals would affect the time series of the polarization angle. There are two important caveats to this strategy. First, periodic variations in polarization measurements may be caused by more prosaic mechanisms, such as seasonal effects at Earth (e.g., Faraday rotation in the ionosphere ~\cite{Yan:2011bq, jackson_classical_1999}), astrophysical phenomena or even instrumental systematic effects. Some of these contributions are, in principle, well known and can be corrected for, like the ionospheric corrections or some of instrumental origin. Others are potentially unknown and would need to be modelled and subtracted before using $\Delta \phi(t)$ effectively to search for DM. Secondly, the analysis depends on unknown \textit{environmental variables} such as $\Delta$ or the stochastic amplitudes $\alpha_i$. This may cause a loss of sensitivity to the ALP birefringence effect in the unlucky situation of near-zero field amplitudes or cancellations in Eq.~\eqref{eq:phi0} during our measurements.

Both problems can be circumvented by performing a global analysis of several sources measured, ideally, at different frequencies and with independent experiments. Using several sources allows one to discard astrophysical periodic effects, as the ALP-induced periodic changes in the polarization of the EM wave should have the same period for all sources, while those of astrophysical origin are expected to have source-specific periods and patterns. On the other hand, different frequencies and experiments allow us to eliminate frequency-dependent changes (e.g. Faraday rotation \cite{jackson_classical_1999}) and instrumental effects, respectively.  Finally, a global analysis enables us to tackle the lack of knowledge about the environmental variables by effectively sampling them statistically from their PDF, such as Eq.~\eqref{eq:Rayleigh} for the stochastic amplitudes.

Before moving to the bulk of the analysis, it is important to stress that the stochastic behavior of the axion field is a \emph{crucial} ingredient to obtain realistic bounds, often overlooked in previous analysis. In the following, to further stress this point, we refer to results both assuming a fixed $\alpha_i=1$ (``deterministic'' scenario, applicable only for $T_{\rm obs} \gg \tau_c$, where the experiment explores the whole distribution of the field values) or by sampling it from the Rayleigh distribution in Eq.~\eqref{eq:Rayleigh} (``stochastic'' scenario, correct for $T_{\rm obs} \ll \tau_c$).

\subsection{Birrefringence searches and UV models of ALP DM}
\label{sec:ALPDMbirefringence}

As discussed in Sec.~\ref{sec:ALPDM} and shown in Eq.~\eqref{eq:ALPDMabundance}, the cosmological abundance of ALP DM is set by its mass and decay constant. The latter is, in turn, related to the axion-photon coupling via an equation of the type
\begin{equation}\label{eq:couplingmodel}
g_{a\gamma} = \kappa\frac{\alpha_{\rm em}}{2\pi}\frac{1}{f_a},
\end{equation}
where $\alpha_{\rm em}$ is the EM fine structure constant and $\kappa$ is a model-dependent coefficient. 
Therefore, once $f_a$ is adjusted for a given mass $m_a$ to provide for the relic DM abundance, one can interpret our search for ALP DM in terms of the sensitivity to different models. In typical QCD-axion models $\kappa$ receives two contributions: an infrared (low-energy) contribution of $\mathcal{O}(1)$ coming from mixing between axion and QCD mesons, and an ultraviolet (high-energy) contribution which usually depends on the electromagnetic and QCD anomaly coefficients. The latter is usually also taken to be of $\mathcal{O}(1)$; however, it can be, in principle, much larger than the infrared contribution, leading to an enhanced axion-photon coupling. This occurs in several scenarios, including the photophilic axion of Ref.~\cite{Farina:2016tgd}, where the axion has an exponentially large coupling
to photons thanks to the “clockwork” mechanism, the generalized Kim-Nilles-Peloso (KNP)
alignment scenarios discussed in Ref.~\cite{Agrawal:2017cmd}, or through a kinetic mixing of abelian
gauge fields, as considered in Ref.~\cite{Daido:2018dmu}, where the axion-photon coupling inherits the (potentially large) dark photon gauge coupling. Therefore, searches of axion DM probing their coupling with photons $g_{a \gamma}$ can always be reinterpreted as measurements or upper limits of the $\kappa$ coefficient (see~\cite{Dror:2020zru} for a related discussion).

\section{Data}
\label{sec:data}

As discussed in Sec.~\ref{sec:ALPbirefringence}, in order to perform a robust search of ultra-light axion DM through its birefringence effects, it is crucial to combine many different sources and, if possible, different experiments observing at different wavelengths. In this work we use an ensemble of observations of the polarization angle of 21 Galactic sources measured by the Parkes Pulsar Time Array (PPTA) at about 1.4 GHz~\cite{Yan:2011bq} and by the QUIJOTE experiment between 10 and 20~GHz with its MFI instrument.

\begin{table}[t]
\centering
\small
\setlength{\tabcolsep}{1.25em}
  \begin{tabular}{cccccc}
  \hline\hline
    Pulsar & $ D_{\rm Earth}$ [pc]  & $D_{\rm GC}$ [kpc]  & $\rho_{\rm DM}$  ${\rm [GeV/cm^3]}$ & OT [yr] & N$_{\rm obs}$ \\
    \hline  
    \texttt{J0437-4715} &  156   &  8.16  & 0.35 $\pm$ 0.06  & 4.40 & 393 \\
    Crab-QUIJOTE        &  2000  & 10.11  & 0.23 $\pm$ 0.04  & 4.56 & 1363 \\
    \texttt{J0613-0200} &  990   &  8.98  & 0.29 $\pm$ 0.05  & 4.71 & 125 \\
    \texttt{J0711-6830} &  110   &  8.11  & 0.35 $\pm$ 0.06  & 4.78 & 47 \\
    \texttt{J1022+1001} &  640   &  8.39  & 0.33 $\pm$ 0.06  & 4.60 & 84 \\
    \texttt{J1024-0719} &  1200  &  8.49  & 0.32 $\pm$ 0.06  & 4.66 & 78 \\
    \texttt{J1045-4509} &  590   &  8.03  & 0.36 $\pm$ 0.07  & 4.78 & 110 \\
    \texttt{J1600-3053} &  1870  &  6.44  & 0.53 $\pm$ 0.09  & 4.78 & 103 \\
    \texttt{J1603-7202} &  3400  &  6.22  & 0.56 $\pm$ 0.10  & 4.66 & 101 \\
    \texttt{J1643-1224} &  1200  &  7.02  & 0.45 $\pm$ 0.08  & 4.60 & 98 \\
    \texttt{J1713+0747} &  1310  &  7.13  & 0.44 $\pm$ 0.08  & 4.78 & 112 \\
    \texttt{J1730-2304} &  470   &  7.67  & 0.39 $\pm$ 0.07  & 4.55 & 70 \\
    \texttt{J1732-5049} &  1875  &  6.45  & 0.52 $\pm$ 0.09  & 3.61 & 22 \\
    \texttt{J1744-1134} &  410   &  7.73  & 0.38 $\pm$ 0.07  & 4.78 & 104 \\
    \texttt{J1824-2452} &  5500  &  2.85  & 1.90 $\pm$ 0.27  & 4.45 & 74 \\
    \texttt{J1857+0943} &  1180  &  7.08  & 0.45 $\pm$ 0.08  & 4.66 & 59 \\
    \texttt{J1909-3744} &  1157  &  7.04  & 0.45 $\pm$ 0.06  & 4.70 & 173 \\
    \texttt{J1939+2134} &  4800  &  6.86  & 0.47 $\pm$ 0.14  & 4.12 & 74 \\
    \texttt{J2124-3358} &  440   &  7.83  & 0.37 $\pm$ 0.07  & 4.78 & 100 \\
    \texttt{J2129-5721} &  7000  &  6.21  & 0.56 $\pm$ 0.10  & 4.56 & 62 \\
    \texttt{J2145-0750} &  710   &  7.79  & 0.38 $\pm$ 0.07  & 4.56 & 67 \\
      \hline\hline
  \end{tabular}
\caption{List of pulsars used in our work including parameters relevant for our analysis: Central value of measured distances to Earth, $ D_{\rm Earth}$, distance to the Galactic center, $D_{\rm GC}$, and local dark matter density, $\rho_{\rm DM}$, calculated with a NFW profile with parameters extracted from \cite{2020Galax837S}, observation time (time difference between last and  first observation), OT, and the number of observations, N$_{\rm obs}$,  for each pulsar used in the analysis. See the main text for a discussion related to the impact in our analysis of the uncertainties in the distance measurements. \label{tab:pulsarlocation}}
\end{table}

In Table~\ref{tab:pulsarlocation} we show the list of the pulsars used in this work and the parameters relevant to our analysis. In particular, knowing the location of the pulsars with respect to Earth and within the Milky Way allows us to estimate the local dark matter density at each source and also determine whether the distance between the source and the observation point is smaller than the coherence length $l_c$. In Table \ref{tab:pulsarlocation} we show the distance of all the pulsars used in our analysis to Earth \cite{Reardon:2021gko,Deller_2008,Manchester_2005, LIGOScientific:2020gml,Kaplan_2008} and to the Galactic center. We also show the corresponding $\rho_{\rm DM}$ at the pulsar estimated using a Navarro-Frenk-White (NFW) profile \cite{1997ApJ490493N} with parameters $\rho_0 = 0.787 \pm 0.037$ GeV cm$^{-3}$ and $R_s = 10.94 \pm 1.05$ kpc, extracted from Ref.~\cite{2020Galax837S}. Note that the pulsar distances to Earth, which are crucial to locate them in the Galaxy, are determined by different methods and in some cases have large uncertainties. Nonetheless, even in the most extreme cases the uncertainty on $\rho_{\rm DM}$, which we report in the Table, is dominated by the NFW parameters uncertainty . Given that the birefringence effect is $\propto\rho_{\rm DM}^{1/2}$, Eq.~\eqref{eq:polALP1}, this uncertainty leads to changes of  order 10\% that we neglect in our analysis.          

Our goal is  to combine long-time measurements of the polarization angle of these pulsars to search for a global modulation $\Delta \phi(t)$  produced by the axion field oscillations, scanning its frequency $\nu_c$ in a certain range that we will determine from the structure of the observation windows below. In the next two subsections we describe the PPTA and QUIJOTE measurements on these sources.

\subsection{PPTA data}
\label{sec:ppta}

The Parkes Pulsar Timing Array (PPTA) started in 2005 and uses the Parkes 64-m radio telescope (ATNF/CSIRO, Australia) to perform continuous observations of a sample of 20~ms pulsars in three radio frequency bands centred at $0.7$~GHz, $1.4$~GHz and $3.1$~GHz, with intervals from two to three weeks, with the primary goal to detect the background of gravitational waves in the pulsar timing band \cite{Manchester_2013}. In this work we use PPTA data released in Ref.~\cite{Yan:2011bq} 
that include polarization measurements of 20 pulsars at 1.4~GHz. These data span a total observation time of $\sim 4.5$ years with a precision that can be as high as $\mathcal O(0.1^{\circ})$ for some pulsars. Further technical details about these observations can be found in~\cite{Yan:2011bq}.

At 1.4~GHz changes in the Faraday rotation arising in the Earth's ionosphere lead to variations of the polarisation angle in scales of one day that are way larger than the data statistical error bar. This effect must be modeled and subtracted from the data \cite{Yan:2011bq}. The PPTA datasets implement two models to treat ionospheric corrections:  the 2007 International Reference Ionosphere (IRI) model-$\texttt{GET}$RM$-\texttt{IONO}$ \cite{Han2006} and the \texttt{FARROT} (Faraday Rotation) software. The latter was developed by  the Dominion Radio Astrophysical Observatory (DRAO), Penticton, Canada~\cite{farrot}. \texttt{FARROT} uses three Chapman layers and a semi-empirical model for variations in these based on the 10.7-cm solar radio
flux measured by DRAO. Instead, IRI uses the model from the International Union of Radio Science (URSI)~\cite{iri}.
\texttt{FARROT} and IRI models incorporate the International Geomagnetic Reference Field2 to take into account the effects from Earth’s magnetic field~\cite{geo}.
These models allow one to study the changes in the rotation measure and in the polarization angle of our signal due to ionospheric variations. In this work we analyze the PPTA data corrected by the IRI model because it gives better results than \texttt{FARROT}~\cite{Yan:2011bq} (see discussion below).

\subsection{QUIJOTE data}
\label{sec:quijote}
PPTA data are supplemented by observations of the Tau A supernova remnant (Crab Nebula) performed with QUIJOTE~\cite{RubinoMartin_2012,Genova-Santos:2015uia} (hereafter Crab-QUIJOTE), an experiment dedicated to observe the cosmic microwave background polarization between 10 and 40 GHz from the Teide Observatory (2400~m a.s.l., Tenerife, Spain).  In particular we use data at 11, 13, 17 and 19 GHz taken with the QUIJOTE Multi-Frequency Instrument (MFI)~\cite{2012SPIE.8452E..33H}, which was operative between 2012 and 2018. Owing to it being the brightest polarised compact source on the sky at the angular resolution of QUIJOTE and in this frequency range, Tau A was the main reference calibrator of the MFI. As such, this source is observed on a regular (virtually daily) basis, using observations in raster-scan mode, lasting $\sim 25$~min to produce maps of size $\sim 8^\circ\times 8^\circ$ around the source to reach deep sensitivities. Tau A was also observed during the QUIJOTE-MFI wide-survey of the full northern sky~\cite{upcomingQUIJOTE1}, using the so-called ``nominal mode''. In this case the telescope is continuously rotated in azimuth at a constant elevation, resulting in a map of a large area of the Northern sky after a full day. Due to its declination ($22.0^\circ$) being very close to Tenerife’s latitude ($28.3^\circ$) Tau A was detected in all these observations. 

\begin{figure}
\begin{center}
\includegraphics[width=1\textwidth]{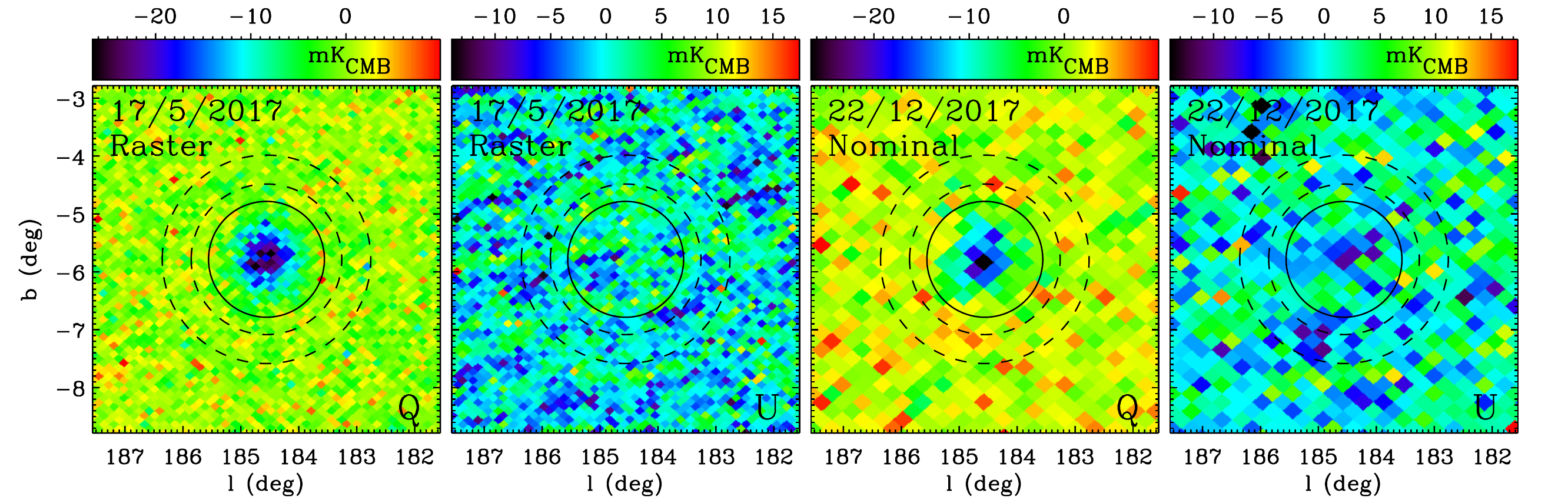}
\caption{QUIJOTE-MFI 11~GHz maps of Stokes Q and U at the position of Tau A, in raster-scan mode (left; performed on 17 May 2017) and in nominal model (right; performed on 22 December 2017), in Galactic projection. The concentric circles depict the regions that are used to derive Q and U flux density estimates (see text for details).}
\label{fig:quijote_maps}
\end{center}
\end{figure}

In Figure~\ref{fig:quijote_maps} we show two examples of QUIJOTE-MFI 11~GHz maps of Stokes Q and U of Tau A, using individual observations in raster and nominal mode. Due to differences in the two scanning strategies, the former leads to a higher integration time per unit area, which results in an improved map sensitivity that is apparent by eye. We gathered a total of 1363 and 685 Tau A observations between November 2012 and December 2018, respectively in these two modes, for a total of 2048 Tau A maps of Stokes Q and U. In each individual map we apply an aperture photometry technique, consisting in integrating the Q and U flux densities enclosed by the solid circle in Figure~\ref{fig:quijote_maps} and subtracting a median background level calculated in the region inside the two concentric dotted circles. The polarisation direction is then obtained as $\gamma=0.5\arctan(-U/Q)$\footnote{The QUIJOTE maps use the COSMO convention, and a minus sign in front of U in order to preserve the same definition of the polarisation angle as in the IAU convention.}. Analogous maps are produced for the other MFI frequencies, so that for each observation we have four estimates of the Tau A polarisation angle at 11, 13, 17 and 19 GHz. MFI has four horns, each of which observe a pair of frequency bands. In this analysis we have used the 11 and 13 GHz frequency bands of horn 3, and the 17 and 19 GHz frequency bands of horns 2 and 4. The 11~GHz provides the best sensitivity, while, due to the combined effect of the fading of the Tau A flux density with frequency (spectral index $\alpha\sim -0.3$) and to the proximity to the atmospheric 22~GHz water vapour line, the signal-to-noise gradually worsens at higher frequencies. We compute for each observation a weighted average of the four polarisation angle estimates, justified by the fact that the ALP birefringence effect is frequency independent (see Sec.~\ref{sec:ALPbirefringence}).We reach sensitivities in the determination of the polarisation angle of $\mathcal O(1-2^{\circ})$ for each observation in raster mode, and a bit worse in the case of the nominal mode. When combining the different frequencies we correct for Faraday rotation of Tau A, using RM$=24.5^\circ/$m$^2$ \cite{weiland11} (this leads to a rotation of the angle of $1.05^\circ$ at the lowest frequency of 11~GHz).

The QUIJOTE-MFI calibration on Tau A consists in setting the global polarization direction, by combining many individual observations. The instrument polarization direction is stable during large periods of time ($\sim 1.5$ years), and therefore variations of the observed polarization angle within these periods could only be due to real on-sky variations. Possible variations in time scales larger than these $\sim 1.5$-year long periods were checked through comparison with the polarization direction measured by WMAP at 22.8~GHz and Planck at 28.4~GHz on the diffuse Galactic emission, and were found to be smaller than $\sim 0.5^\circ$. Further details on this study will be provided in different upcoming papers (\cite{upcomingQUIJOTE1} and \cite{upcomingQUIJOTE2}).

The QUIJOTE-MFI instrument underwent different technical modifications during its 6-year lifetime, leading to differences in the effective instrument sensitivity as a function of time. In this study we decided to remove data taken before April 2014, which was less precise due to the lack of a reference calibration diode, and  we end up with a sample of 1099 QUIJOTE-MFI estimates of Tau A polarisation direction, spanning a period of 4.5 years.

\section{Data analysis and results}
\label{sec:analysis}
\subsection{Generalized Lomb-Scargle periodograms}
\label{sec:methodLS}

The best approach for our endeavor and, in general, to search for periodic signals in unevenly distributed time series is the Lomb-Scargle (LS) periodogram~\cite{1982ApJ...263..835S,1976Ap&SS..39..447L}. This is a standard tool in astronomy based on a combination of techniques of signal analysis: Fourier methods, least squares, phase-folding, and Bayesian approaches~\cite{2018ApJS..236...16V}. The central element of the method, the LS periodogram, which we label $P_{LS}(\nu)$ in the following, can be regarded as the analog of the power spectrum in continuous Fourier analysis (see Refs.~\cite{2018ApJS..236...16V,Vio2018} for subtleties of this interpretation). It can also be understood as the result of a least-square minimization process, which can be extended to include the uncertainties of the data. This leads to the \textit{generalized} LS periodogram~\cite{Zechmeister:2009js,2018ApJS..236...16V}, which is the version that we use in our work. In Appendix~\ref{sec:periodogram} we provide a brief description of the generalized LS periodogram, the standard procedure we follow to calculate it and its relation to a statistical minimization procedure.

The significance of any peak appearing in $P_{LS}(\nu)$ can be assessed by a False Alarm Probability (FAP). The FAP quantifies the probability that a data set with no periodic signal leads to such peak by coincidental alignment of random fluctuations~\cite{2018ApJS..236...16V}. 
The most robust method to compute the FAP is the \textit{bootstrap approach}, which resamples the time series by keeping the temporal coordinates but with observations randomly distributed using Monte-Carlo (MC) simulations \cite{VanderPlas_2015}.
The maximum peaks obtained in these MC simulations form a distribution, which can be used to directly compute the FAP.

To apply the LS technique and extract reliable bounds, it is important to consider  the frequency limits and the grid spacing of our data~\cite{2018ApJS..236...16V}. For a pulsar observation spanning a total time $T_{\rm obs}$, a signal with frequency $1/T_{\rm obs}$ will complete exactly one oscillation cycle, providing a suitable minimum frequency $\nu_{\rm min}$ for the LS periodogram\footnote{For lower frequencies, the periodic harmonic signal can be confused with a polynomial fit.}. For the maximum frequency $\nu_{\rm max}$ we choose the inverse of the minimum separation between observations that we identify with the pseudo-Nyquist frequency (see~\cite{2018ApJS..236...16V} for more details). Taking as a reference the observations of \texttt{J0437-4715} by PPTA~\cite{Yan:2011bq}, $\nu_{\rm max}\simeq20$ days$^{-1}$ is determined by the minimum separation between observations of $\sim$ 1 hour and $\nu_{\rm min}\simeq6\times 10^{-4}$ days$^{-1}$ by $T_{\rm obs}\sim 4.4$ yrs. As a result, the range of ALP masses that can be probed spans about four orders of magnitude, $10^{-23}\lesssim m_{a}\lesssim 10^{-19}$ eV. 

Once the frequency interval is identified, one must choose a frequency resolution. The grid must have enough resolution to avoid the possibility of losing peak power. However,   a large number of points may make  the search for peaks--and hence the production of bootstraps--prohibitive in computational cost. To ensure a frequency grid of good resolution for each peak, we over-sample by a factor $n_{0}=5$ per peak, and   use a grid of spacing $\nu_{\rm min}/n_{0}$. With this choice, the total number of required periodogram evaluations is $N=n_{0}T_{\rm obs}\nu_{\rm max}$~\cite{VanderPlas_2015, Richards_2012}. 

\subsection{Searching for peaks in the power spectrum}
\label{sec:peakhunting}

\begin{figure}
\begin{center}
\begin{tabular}{cc}
\hspace{-0.5cm}
\includegraphics[width=0.55\textwidth]{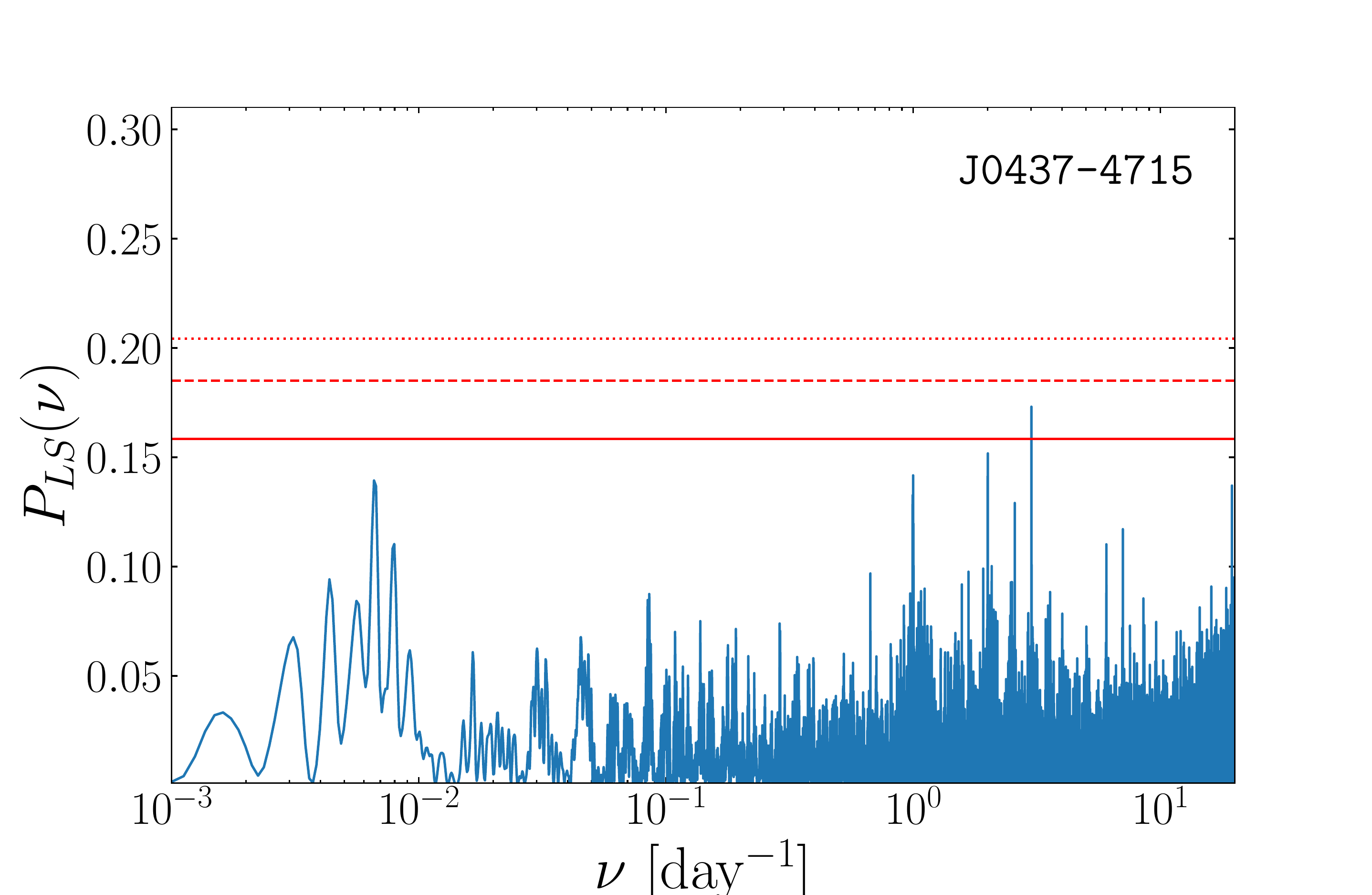}
\hspace{-1.1cm}
\includegraphics[width=0.55\textwidth]{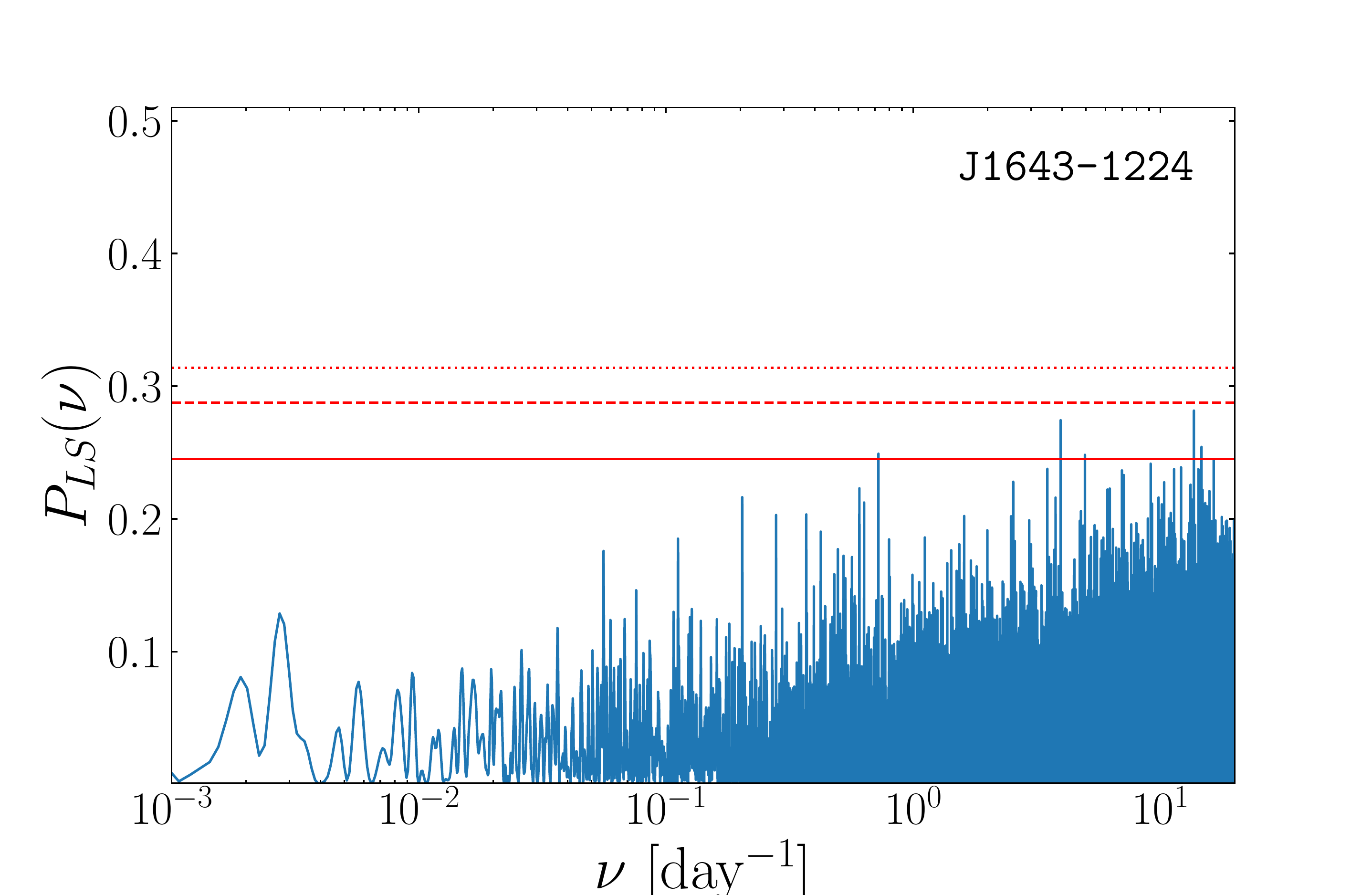}\\
\hspace{-0.5cm}
\includegraphics[width=0.55\textwidth]{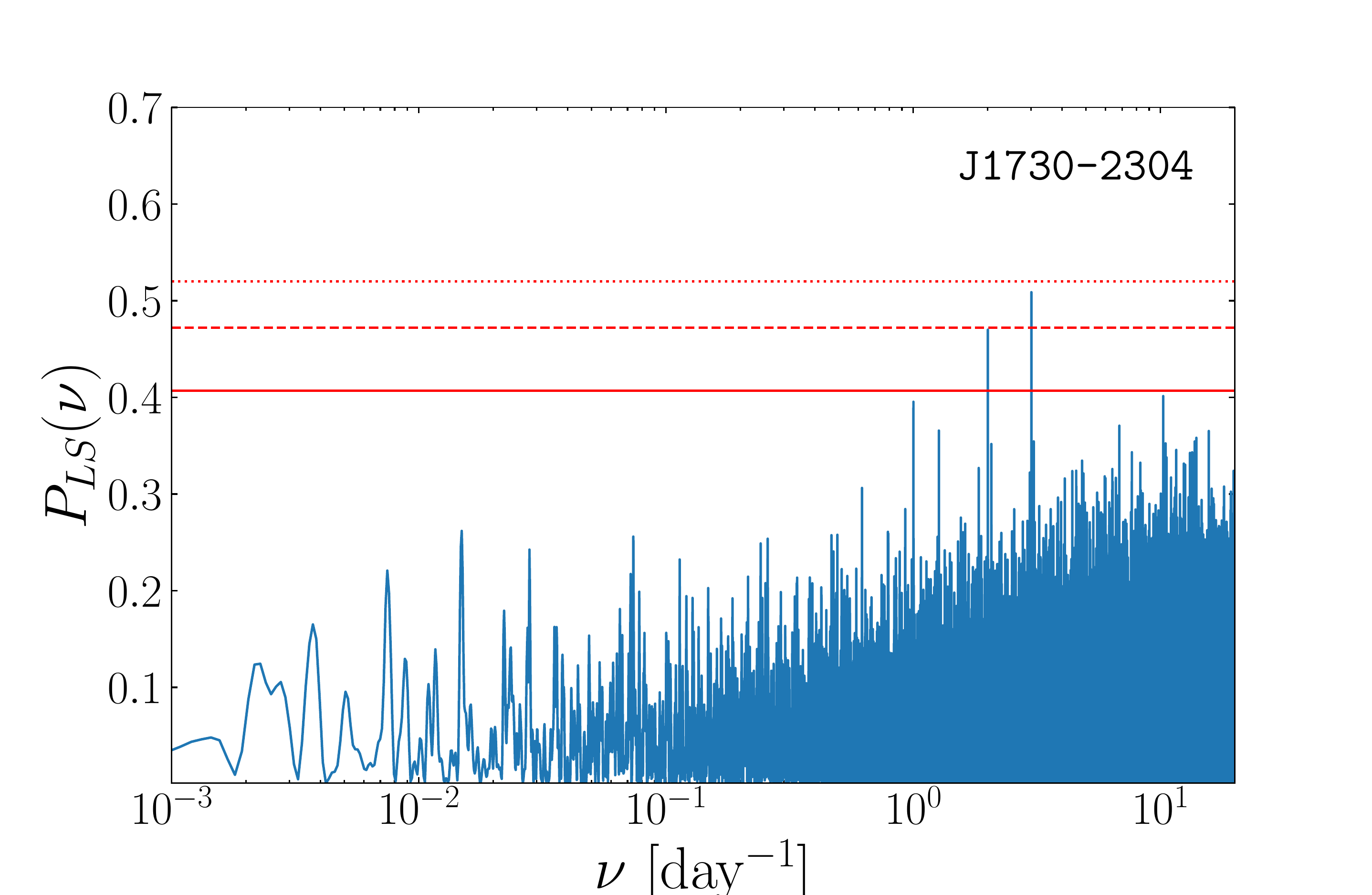}
\hspace{-1.1cm}
\includegraphics[width=0.55\textwidth]{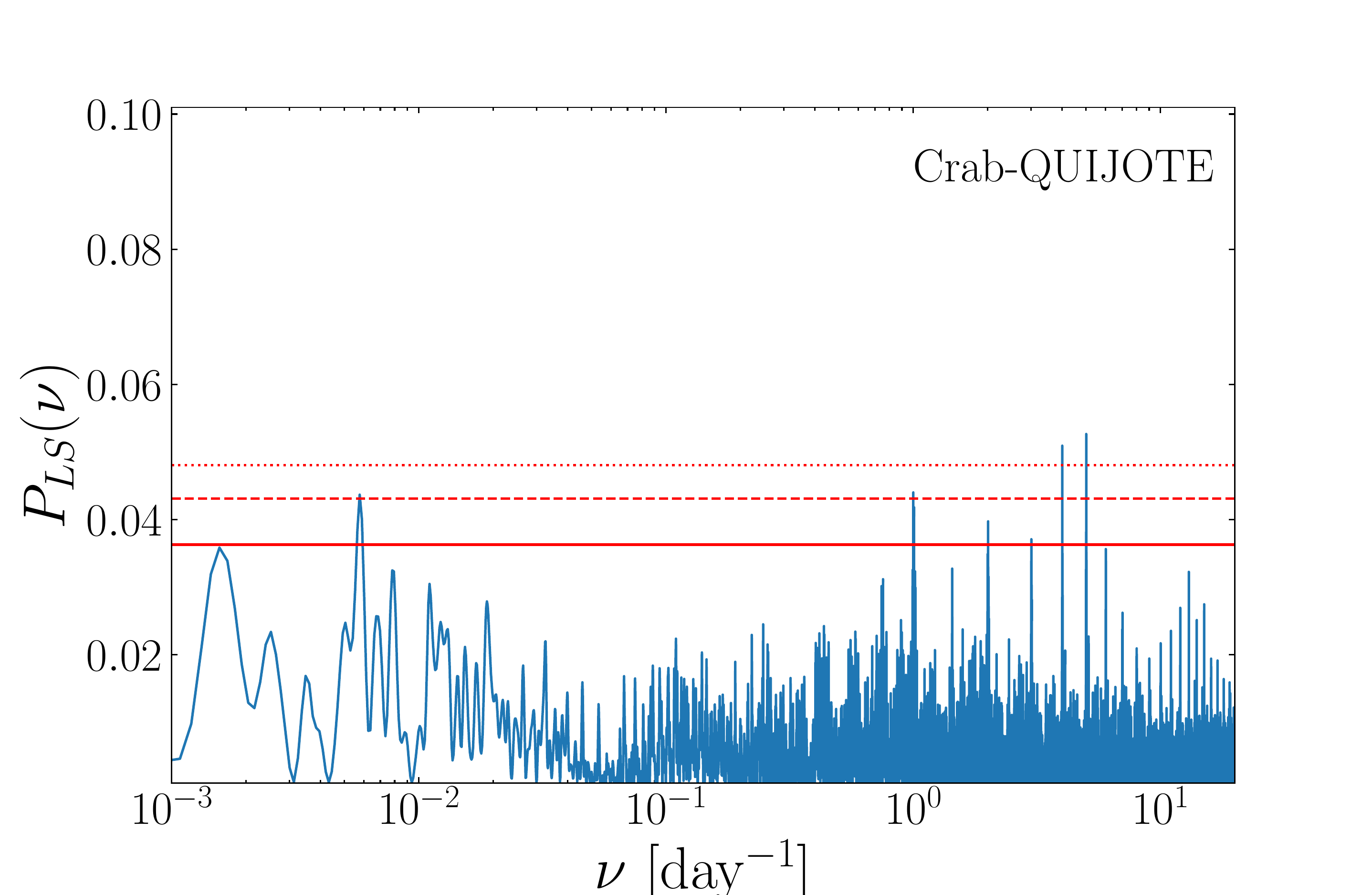}
\end{tabular}
\caption{Generalized LS periodograms for the time series of polarization measurements for four of the pulsars in our analysis and including the IRI ionospheric corrections in the case of PPTA observations. 
We also show FAPs at 32\% (solid line), 5\% (dashed line) and 1\% (dotted line) false-positive rates estimated using a bootstrap method with a 1000 random resamplings of the data set at the same temporal coordinates.}
\label{fig:LSperiodograms}
\end{center}
\end{figure}

After adjusting the range and density of the grids, we compute the generalized LS periodograms corresponding to the polarization time series of the 20 PPTA pulsars (implementing the IRI ionospheric corrections) and of Crab-QUIJOTE. In each case we also compute the FAPs with false-probability rates of 32\%, 5\% and 1\%,  using 1000 MC bootstraps of the data. For the PPTA pulsars we observe that the highest values of $P_{LS}(\nu)$, and those of the FAPs, are lower when the data set has small individual errors and dispersion (e.g., see the time series $\Delta\hspace{0.05cm}\text{PA}$ for \texttt{J0437-4715} source in Fig. \ref{Fig:TimeSeries}). In general, we do not find peaks with FAPs smaller than 1\% (or even smaller than 5\% in most cases). There are two exceptions: \texttt{J1730-2304} and Crab-QUIJOTE, for which we find peaks with false-rate probability similar or smaller than 1\%. This is illustrated in  Fig.~\ref{fig:LSperiodograms}, where we show the periodograms for two of the statiscally most powerful\footnote{In the sense of smallest nominal errors and dispersion of the data around the mean.} PPTA pulsars, \texttt{J0437-4715} and \texttt{J1643-1224}, and for the two pulsars where prominent peaks appear. It is important to note that the periodograms of the time series \textit{without} ionospheric corrections show prominent peaks at $\nu\sim1$ day$^{-1}$ and $\nu\sim1$ yr$^{-1}$ for almost all the PPTA pulsars. This is precisely the type of seasonal pattern that is expected to be produced by ionospheric effects. Therefore, subtracting them is important for our analysis, albeit at the cost of introducing some dependence on the corresponding modelling. In particular, the peak found in \texttt{J1730-2304} seems to be spuriously introduced by these models as it does not appear in the uncorrected data (see Appendix ~\ref{sec:ionosph} for details about a LS analysis of these ionospheric corrections). In the case of Crab-QUIJOTE, we observe 
two peaks surpassing FAPs significance of 1$\%$. This indicates the presence of periodic signals, the most significant of which corresponds to $\sim5$ day$^{-1}$. 

The absence of similar peaks in the power-spectrum of all the other pulsars, especially in the statistically most powerful ones, such as \texttt{J0437-4715} and \texttt{J1643-1224}, suggests that the peaks observed with low FAP are not produced by the oscillations of the ALP field. In order to address this question precisely we first need to translate the absence of peaks for a given source into an upper limit on the value of the amplitude of the oscillation. Finally, comparing and combining different pulsars requires expressing this bound as an upper limit on the axion-photon coupling $g_{a\gamma}$.

\subsection{Constraints from single sources}
\label{sec:individual}

Next, we perform a new set of MC simulations generating 3000 pseudo-experiments, as done in the bootstrap method for the calculation of the FAPs, but injecting a harmonic oscillation in the time series
\begin{equation}
    \Delta \phi_{\rm{sim}} = \phi \cos\left(2\pi\nu\, t + \varphi\right),
\end{equation}
with a given frequency $\nu$ and amplitude $\phi$, and a random phase $\varphi$ sampled from $[0,2\pi]$ with an uniform PDF. Periodicity for the polarization angle is incorporated through resamplings of the time series and thus in the LS periodograms using constraints over $\phi$ in the simulated values of $\Delta\phi_{\rm{sim}}$.
With the population of pseudo-experiments, we compute the PDF of the corresponding LS periodograms $P_{LS}(\nu,\phi)$ and find the value corresponding to the lower 5\% tail of the distribution (i.e., such that the probability of obtaining a $P_{LS}(\nu,\phi)$ larger than that value is 95\%) that we call $\bar P_{LS}(\nu,\phi)$.  We then find numerically the value $\phi_{95}$ for which  $\bar P_{LS}(\nu,\phi_{95})$ is equal to the experimental $P_{LS}(\nu)$. Hence, $\phi_{95}$  can be interpreted as an upper bound on $\phi$ at 95\% confidence level for the given frequency $\nu$\footnote{Important aspects of this method 
are similar to the techniques used to detect tertiary companions-exoplanets in binary systems, see Ref.~\cite{Muterspaugh_2010}.}. To smear out the fluctuations related to the MC realizations over the LS periodograms, we smooth out the limits considering segmentation of the bound using a rolling mean method\footnote{In the context of nonlinear signal processing,  Ref.~\cite{arce_2005} gives a general discussion of the rolling mean method (RMM). On the other hand, it is also possible to relate the fluctuations to statistical variances of the periodograms using Barttlet's method~\cite{Bartlett1950}, which gives a smoothing of the LS periodograms through averaging with some filters over the neighbouring power spectrum frequencies ending in a method similar to RMM. In MC-LS periodograms lower points arise in different frequencies (windows) yielding stronger bounds of $\phi_{95}$, hence the most accurate method to treat the under-fluctuations is one with dynamical windows as given by the RMM.}: at a given frequency $\nu_{j}$, the new output for $\phi_{95}(\nu_{j})$ is the average of the $\phi_{95}$ within a window centered at $\nu_{j}$, where we take as the reference the $\phi_{95}$ at the frequencies $\nu_{j-1}$ and $\nu_{j+1}$. Now the output for $\phi_{95}(\nu_{j+1})$ is the average of the $\phi_{95}(\nu_{j})$ and $\phi_{95}(\nu_{j+2})$ within the window centered in $\nu_{j+1}$. This procedure is repeated for all $j$ with the same window size for all grid frequencies and all pulsars.  Thus, at each frequency $\nu_{j}$, the rolling mean computes a new estimate for $\phi_{95}$.

\begin{figure}[t]
\begin{center}
\begin{tabular}{cc}
\includegraphics[width=0.85\textwidth]{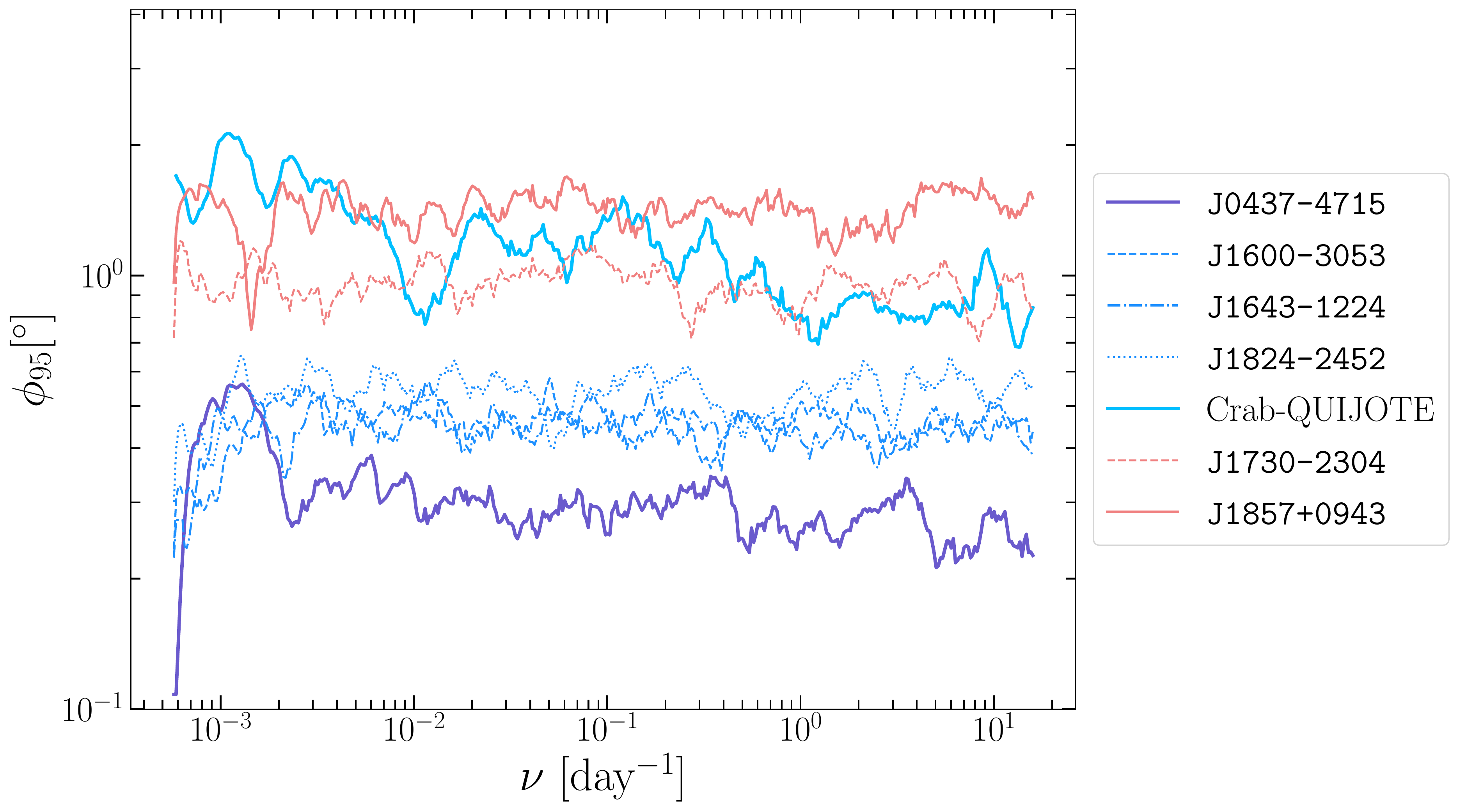} 
\end{tabular}
\caption{Bounds for $\phi_{95}$ for four competitive PPTA pulsars (violet and blue) with the strongest limits. In addition, we consider two marginal PPTA pulsars (red) with the weakest limits. Remaining pulsars from PPTA have bounds among these extreme cases.  We also show the results of the Crab pulsar (cyan).}
\label{fig:individual}
\end{center}
\end{figure} 

Figure \ref{fig:individual} shows individual $\phi_{95}$ bounds for seven pulsars from Table \ref{tab:pulsarlocation}, corresponding to the four PPTA pulsars with the most competitive bounds (\texttt{J0437-4715}, \texttt{J1600-3053}, \texttt{J1643-1224}, and \texttt{J1824-2452}), two PPTA pulsars with the weakest limits (\texttt{J1024-0719}, \texttt{J1730-2304}) and Crab. At low frequencies, the pulsar \texttt{J1600-3053} gives the strongest bound in the range of $6\times10^{-4}$ day$^{-1}\lesssim\nu\lesssim10^{-3}$ day$^{-1}$. Instead, for frequencies greater than $10^{-3}$ days$^{-1}$, the pulsar \texttt{J0437-4715} bounds dominate over those of the other pulsars, giving the strongest constraint on $\phi_{95}$. The rest of pulsars from PPTA have constraints among these limits. Finally, the analysis of Crab-QUIJOTE leads to bounds in the ballpark of those from the weakest PPTA pulsars. Nonetheless, the difference of the bounds obtained from the statistically stronger and weaker pulsars is not dramatic and is generally not larger than a factor $\sim5$.

We emphasize at this point that, for each pulsar, $\phi_{95}$ is just an experimental limit on the amplitude of a harmonic signal that is assumed to appear overlaid on a random resampling of the data. An upper limit on the axion-photon coupling $g_{a\gamma}$ can be obtained by assuming that this harmonic signal is entirely produced by the ALP DM field oscillations; namely, $\phi_a=\phi$ in Eq.~\eqref{eq:polALP2}. This is a conservative assumption at the level of setting constraints, as we are allowing the ALP to saturate any possible periodic signal in the data. We can then use Eq.~\eqref{eq:phi0} to extract a bound on $g_{a\gamma}$ from  the one on $\phi_a$. 
The amplitude $\phi_a$ is a random variable whose PDF can be extracted from the MC studies and the data as described above.  For simplicity, we model this PDF for each source and frequency as a Rayleigh distribution with a scale parameter $\sigma(\nu)$ adjusted to match optimally the PDF obtained directly from MCs\footnote{This model works quite accurately in all the cases in which we have tested it and is only improved marginally when using more complex distributions, such as the Rice or Weibull distributions, featuring additional parameters with the Rayleigh distribution as a limiting case~\cite{Stats}.}. The PDF of $g_{a\gamma}$ then follows from the one of $\phi_a$ and from those of the environmental variables: $\Delta$ in the deterministic case or $\Delta$, $\alpha_o$ and $\alpha_s$ in the stochastic case. Surprisingly, we find that in our setup the sensitivity to $g_{a\gamma}$ improves in the stochastic analysis compared to the deterministic one. We refer the reader to Appendix~\ref{sec:AppVs}, where we perform a toy MC study of this comparison. In the following, we derive the bounds on $g_{a\gamma}$ using the stochastic analysis. 

For each PPTA pulsar and Crab-QUIJOTE, we generate 10$^5$ MC {\it pseudo-experiments}. In each one we draw a measurement of $\phi_a$, with the corresponding distributions adjusted above, and a value of the environmental variables $\Delta$, $\alpha_o$ and $\alpha_s$. Here we assume that the source and Earth are in different spatial coherence patches ($D_{\rm Earth}\gg l_c$). By comparing the data in Tab.~\ref{tab:pulsarlocation} and Eq.~\eqref{eq:coherenced}, one can see that this assumption is valid for all pulsars and the whole mass range except for \texttt{J0437-4715} and \texttt{J0711-6830} in the mass region close to the lower end $m_{a}\simeq 3\times 10^{-23}$ eV. Given that these pulsars do not lead to the strongest constraints in that region, see Fig.~\ref{fig:individual}, we neglect the effect of spatial Earth-source coherence in our analysis. Also notice that the physical extent of the Crab-remnant of $\sim$3.5 pc is smaller than $l_{c}$ for all masses $m_{a}\lesssim 2\times10^{-21}$ eV (with typical dispersion velocities $\sigma\sim10^{-3}$). In this regime, we can consider any point in the remnant as a source of the same ALP spatially-coherent field.

Using Eq.~\eqref{eq:phi0} and the values of $\rho_{\rm DM}$ displayed in Tab.~\ref{tab:pulsarlocation}, we obtain the individual PDFs of $g_{a\gamma}$ as a function of $m_a$.
The strongest bounds are derived from the four statistically most powerful PPTA pulsars, \texttt{J0437-4715}, \texttt{J1600-3053}, \texttt{J1643-1224}, and \texttt{J1824-2452} shown in Fig.~\ref{fig:individual}. The relatively weaker bound derived on $\phi_a$ from \texttt{J1824-2452} is compensated by the higher $\rho_{\rm DM}$ density estimated at source. In the end, these four pulsars lead to relatively similar bounds in all the considered mass range. For instance, at $m_a=10^{-22}$ eV one obtains $g_{a\gamma}\lesssim10^{-12}$ GeV$^{-1}$ at 95\% C.L., which is in the same ballpark as the limits derived from other astrophysical probes (see below).   

Finally, note that the discussion here is purely statistical and assumes not only that we can repeat the experiments for a given pulsar ($\phi_a$), but also that we can draw different values for the environmental variables $\Delta$ and the $\alpha_i$. Since these are unknown but fixed parameters, the only way to sample their distributions in our given Universe is by performing a global analysis including many sources, as done below.

\subsection{Constraint from a combined analysis}

\begin{figure}[t!]
\begin{center}
\includegraphics[width=0.75\textwidth]{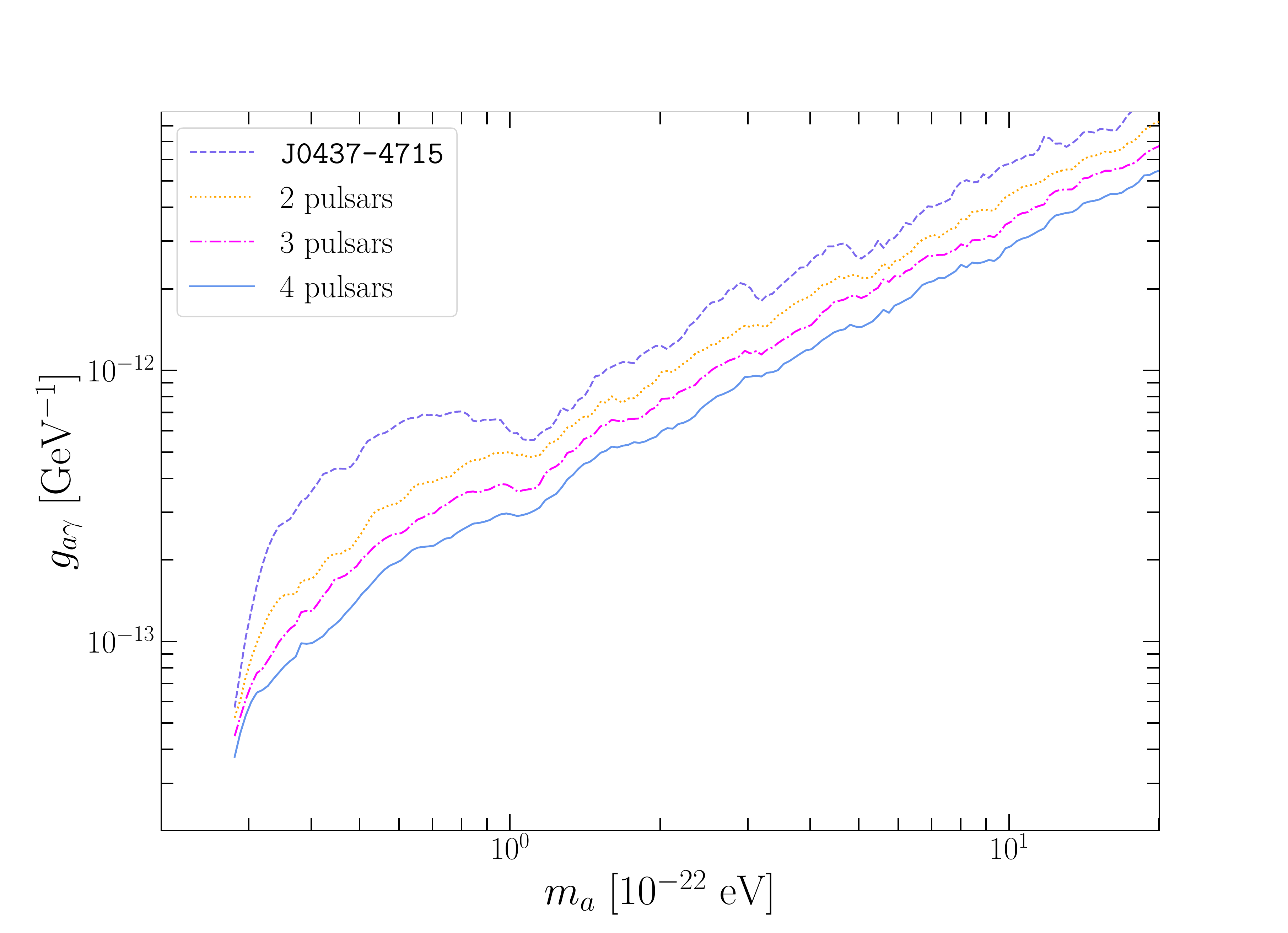} 
\caption{Effect of combination of different pulsars for the stochastic case. We start with one source here: \texttt{J0437-4715}. The remaining pulsars considered are \texttt{J1600-3053}, \texttt{J1643-1224}, and \texttt{J1824-2452}.}
\label{fig:constraints0}
\end{center}
\end{figure}

We are now ready to derive the combined bounds from the PPTA and Crab-QUIJOTE pulsars. In order to do this we carry out a MC analysis, as in the previous section. In each pseudo-experiment (or MC run) we obtain a value of $g_{a\gamma}$ for each PPTA pulsar and Crab-QUIJOTE. The global value $g_{a\gamma}$ is obtained as a weighted mean. By running several pseudo-experiments, we are able to construct its PDF.

To be more precise: Under the Rayleigh model for the distributions of $\phi_a$ obtained in the previous subsection, we compute the scale $\sigma_{i}(\nu_{j})$ for each pulsar ($i=1,...,N_{\text{pulsars}}$) and each frequency we are probing. We then generate 10$^5$ MC pseudo-experiments drawing a value of $\phi_{a,i}$ for each pulsar. We also draw a value of the environmental variables $\alpha_o$ (for the ALP amplitude at Earth) and the $\alpha_{s,i}$ and $\Delta_i$ for the pulsars. Using Eq.~(\ref{eq:phi0}) we obtain the values of $g_{a\gamma,i}$ for each pulsar.  
The weights for the weighted mean of the $g_{a\gamma}$ (or global average) are found by propagating the root squares of the variances of $\phi_{a,i}$, which are directly related to the scale factor in the Rayleigh distribution. We report our upper limit on the axion-photon coupling as the 95\% C.L. value obtained from the resulting distribution of the global $g_{a\gamma}$.

As an illustration of the impact of combining several pulsars, in Fig.~\ref{fig:constraints0} we show the combination of the 4 pulsars with the strongest \emph{individual} $\phi_a$ bounds (see Fig.~\ref{fig:individual}). We observe that as more pulsars of similar sensitivity are added into the combined analysis, the reach of the bound increases. The pulsar \texttt{J1824-2452} contributes in a significant way to the combination due to  its higher $\rho_{\rm DM}$, despite not being as statistically powerful as the others. 

We are also now in the position to analyse the peaks in the periodogram discussed in Sec.~\ref{sec:peakhunting}, appearing in \texttt{J1730-2304} and Crab-QUIJOTE with FAPs overpassing (or close to) 1\% significance. In particular, for a given peak we compare the PDF of $g_{a\gamma}$ at the given frequency and pulsar with the global PDF of $g_{a\gamma}$ at that same frequency and obtained excluding that same pulsar. In so doing, we find that the hypothesis that the peak observed by \texttt{J1730-2304} at $\nu\simeq3$ days$^{-1}$ or by Crab-QUIJOTE at $\nu\simeq5$ days$^{-1}$ is produced by the interactions of the photons with the ALP field is excluded by the global analysis in both cases with a significance larger than 99.999\% \footnote{More precisely, the value that we are excluding is the median of the PDF.}.   
 
\begin{figure}[t!]
\begin{center}
\begin{tabular}{cc}
\includegraphics[width=0.75\textwidth]{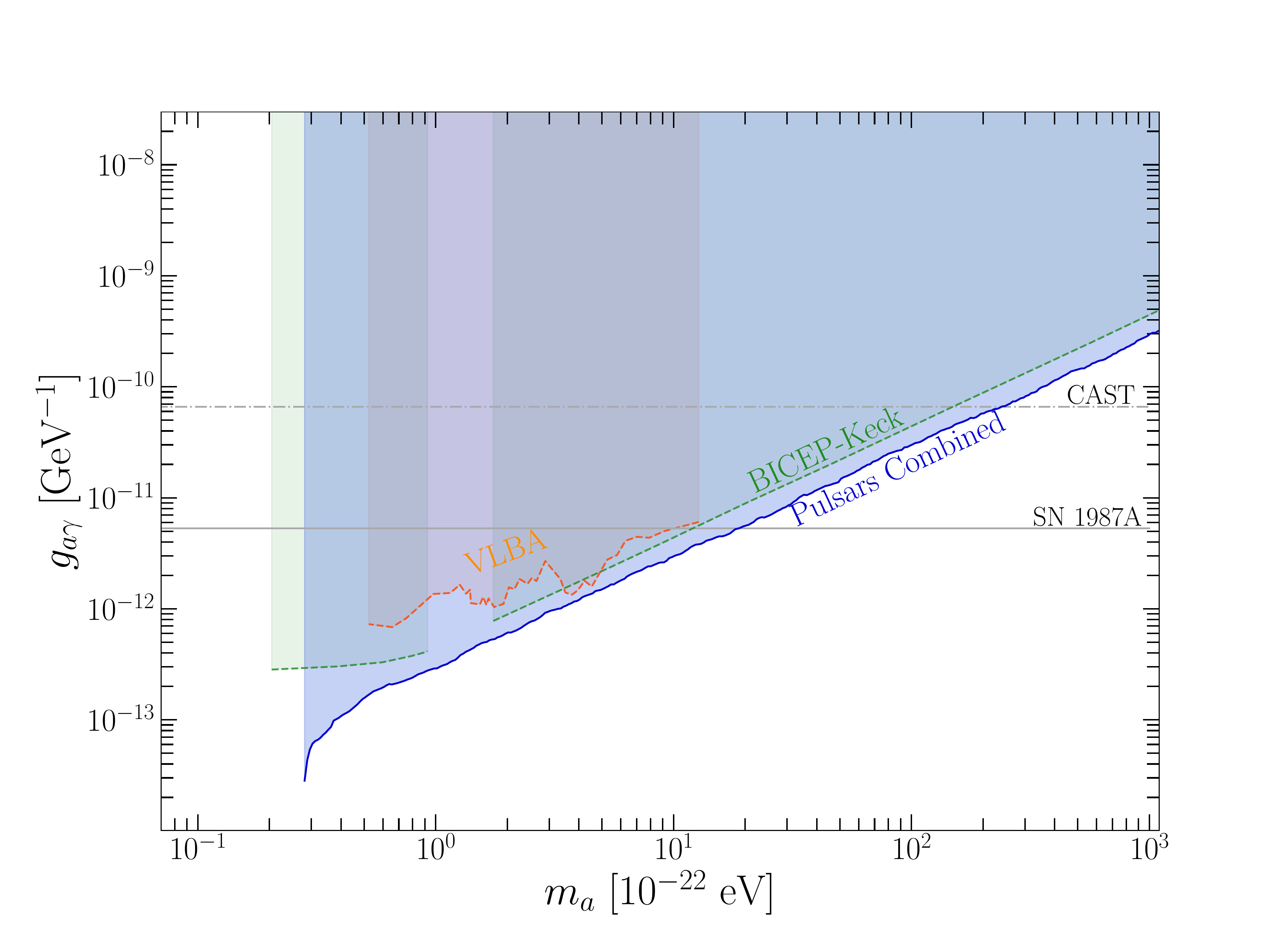} 
\end{tabular}
\caption{Bounds on the axion-photon coupling as a function of the ALP mass. The blue area indicates the excluded region at 95\% C.L. obtained in this work from the combination of PPTA and Crab-QUIJOTE pulsars including stochastic effects.  The darker gray dot-dashed line indicates the region excluded by CAST experiment~\cite{Cast2017}. The gray solid line is the limit from supernova SN~1987A~\cite{Payez:2014xsa}. The orange area indicates the region excluded by MOJAVE VLBA. The green region is the excluded zone by BICEP-Keck (smoothed bound) \cite{keckcollaboration2021bicep}. }
\label{fig:constraints1}
\end{center}
\end{figure}

In Fig.~\ref{fig:constraints1} we show the final bound including all 20 pulsars from PPTA and Crab-QUIJOTE. The 95\% C.L. upper combined limit for the ALP coupling is comparable to other bounds obtained by searching for the same effect either in parsec-scale jets from active galaxy nuclei by MOJAVE VLBA~\cite{Ivanov:2018byi} or from CMB by BICEP-Keck~\cite{keckcollaboration2021bicep} (for the origin of these acronyms see the corresponding references). 
The bound on $g_{a\gamma}$ derived in this paper is the strongest for all the masses in the range $3\times 10^{-23}\text{ eV}\leq m_a\leq2\times 10^{-21}\text{ eV}$. In particular, in the mass regions around $m_a\simeq3\times 10^{-23}\text{ eV}$ and $m_a\simeq10^{-22}\text{ eV}$, our limit is stronger than the other ones by a factor $\sim 5$. For masses approximately above $2\times 10^{-21}\ $eV the strongest limit is set by supernova SN~1987A cooling via Primakoff effect, via a null search of $\gamma$ rays~\cite{Payez:2014xsa}. The strongest limit by a direct experimental search is set by the CAST experiment, by searching for axions emitted by the Sun~\cite{Cast2017}. 

It is important to stress that neither the analysis of VLBA-MOJAVE nor BICEP-Keck incorporate the stochastic nature of the ULDM field amplitude. As we have discussed in our analysis, this might have a significant impact in the quoted upper limits. Finally, we also note that our bounds are considerably weaker (and extend over a shorter range of masses) as compared to those derived from \texttt{J0437-4715} in~\cite{Caputo:2019tms} by some of the authors of this paper, using the same dataset. The main reasons for this difference are: \textit{(i)} the robust statistical treatment provided by the FAPs in a periodogram analysis of time series with large dispersion of central values and small nominal errors, in contrast to a standard frequentist analysis; and \textit{(ii)} a proper treatment of the environmental variables in a global analysis of data from different pulsars. In Appendix~\ref{sec:comparetofreq} we present a more detailed discussion of this comparison. In summary, the  limits from the PPTA and Crab-QUIJOTE pulsars on the axion-photon couplings presented in this work update and improve (in terms of quality and robustness of analysis) those presented in~\cite{Caputo:2019tms}.   
  
\section{Future prospects}
\label{sec:forecasts}
\begin{figure}[t]
\begin{center}
\begin{tabular}{cc}
\includegraphics[width=0.75\textwidth]{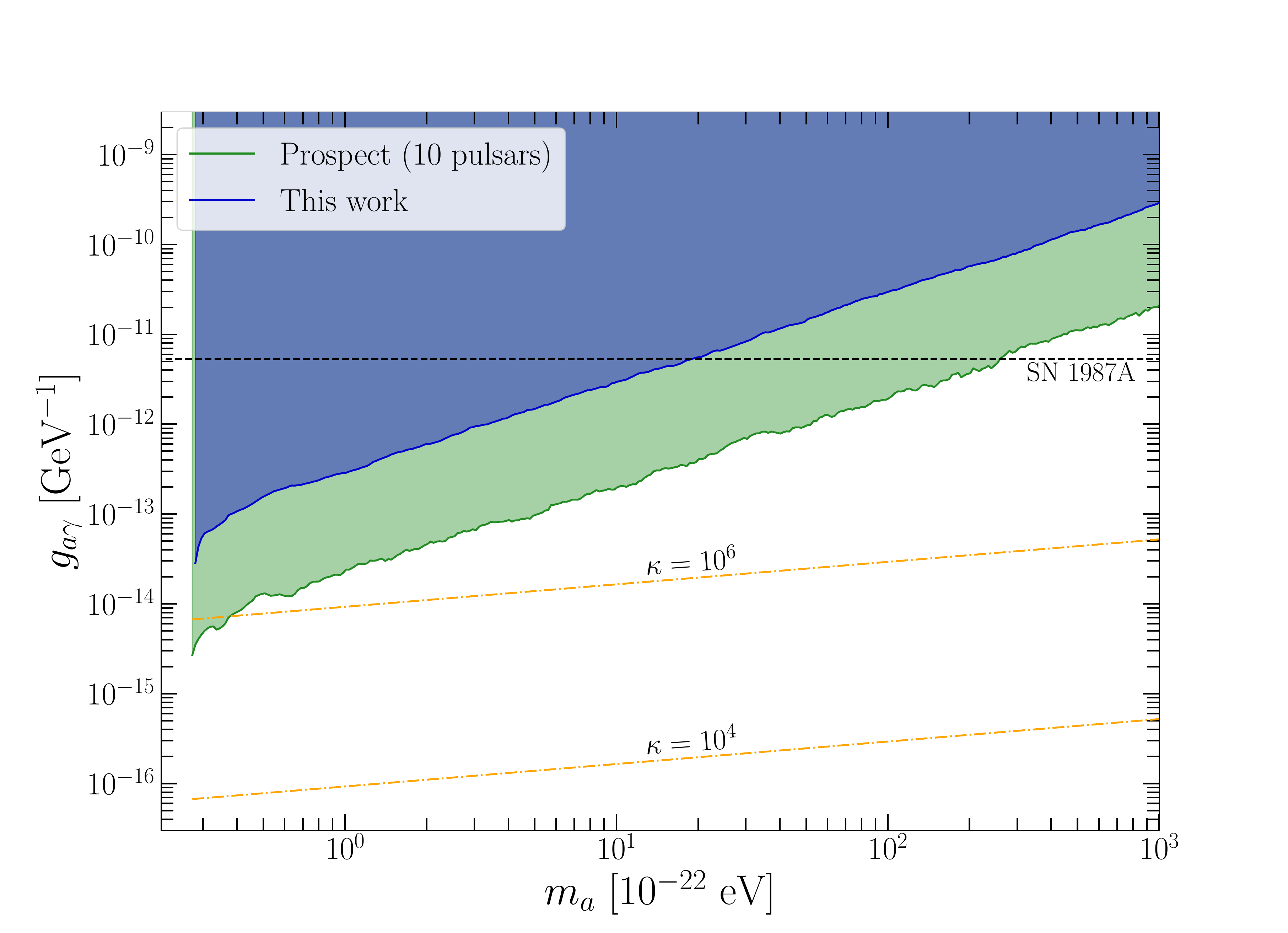} 
\end{tabular}
\caption{
Bounds on the axion-photon coupling as a function of the axion mass at 95$\%$ C.L. for 20 Galactic pulsars, as presented in Fig. \ref{fig:constraints1} (blue), and for the future prospect of observing 10 pulsars with similar precision to \texttt{J0437-4715} during a span of 5 years located 
closer to the Galactic center with $\rho_{\rm DM}=46.75$ GeV/cm$^3$ 
(green). The black dashed line is the limit from supernova SN~1987A~\cite{Payez:2014xsa}. In orange the values of $g_{a \gamma}$ for an axion that makes up 100\% of the DM density following Eqs.~\eqref{eq:ALPDMabundance} and \eqref{eq:couplingmodel}, for a fixed value of $\theta_0=1$. }
\label{fig:prospects}
\end{center}
\end{figure}

Let us finally comment on future directions and possible ways in which our analysis can be further improved. One simple extension would be to consider data from pulsars closer to the Galactic center region 
One of the problems with this strategy is that the presence of hot, ionized gas in the central part of our Galaxy may lead to a decrease in the observed flux density from neutron stars residing in this region. Nevertheless, future surveys such as the  Square Kilometre Array (SKA)~\cite{SKA-JapanPulsarScienceWorkingGroup:2016tcw} and Next Generation Very Large Array (ngVLA)~\cite{carilli2015generation} are expected to probe a sizable population of pulsars in the Galactic center. This should provide better constraints because of the larger DM density in those regions (the limit on $g_{a \gamma}$ scales linearly with the square root of the density), and also robust ones because of the large number of expected detections. In Fig.~\ref{fig:prospects} we show a forecast of the potential reach of an analysis of 10 pulsars, each including 500 observations spanning 5 years with a dispersion and precision similar to the  \texttt{J0437-4715} data, and with $\rho_{\rm DM}=46.75$ GeV/cm$^3$. 

Another intriguing possibility is to consider polarization data from pulsars hosted by globular clusters. At the moment there is no consensus on the DM profiles of globular clusters, see e.g.~\cite{Ibata:2012eq, Bradford:2011aq, Mashchenko:2004hj}. Nevertheless, it is plausible that order $1\%$ of their content is still made of DM today~\cite{Gao:2004au}, after being drastically reduced because of tidal stripping and dark matter thermalization~\cite{Saitoh:2005tt}\footnote{More aggressive scenarios assume more significant profiles for some globular clusters, see e.g.~\cite{McCullough:2010ai,Fortes:2019gfe}.}~. Even if only $\mathcal{O}(1)\%$ of the total amount of matter in globular clusters is dark matter, its density in the inner cores may be $\mathcal{O}(10^2-10^3)$ larger than the local one. This would imply again a sensitivity to an ALP-photon coupling $10-30$ times smaller than the one probed in the present work, assuming the same precision on the measurements. Furthermore, globular clusters provide a natural environment for pulsars to form, see \cite{GCpsr}. In both cases, it is very likely that future bounds will reduce the allowed parameter space by more than an order of magnitude in the coupling $g_{a\gamma}$, an achievement hard to foreseen with any other probes for this range of masses.  

Finally, another line of improvement could be achieved by using correlations between different pulsar mesured with Pulsar Polarisation Arrays as proposed in \cite{Liu:2021zlt}.

\section{Conclusions}

We have studied the effects of ULDM on the \emph{polarized }radio waves that we detect from pulsars. We focused on the coupling $g_{a\gamma}$ in \eqref{eq:Lagrangian}, which makes the ULDM  behave as a chiral medium harmonically affecting the polarization angle of photons while they are in transit from the astronomical source to Earth. Our work improves on previous analyses in several ways. First, we use techniques based on periodograms to analyze the time series of the polarization measurements. The calculation of the false-alarm probability rates, using bootstrap (Montecarlo sampling) techniques, provide us with robust upper limits on the strength of harmonic signals in the data. Second, we include systematically the environmental variables in the analysis, see Eq.~\eqref{eq:phi0}. In particular, we carefully considered  the intrinsic degree of stochasticity of this signal, arising from the virialized state of ULDM in the Milky Way's dark matter halo. At the practical level, this transforms the  local energy density $\rho_{\rm DM}$ into a random variable Rayleigh-distributed with scale parameter given by the standard value in WIMP DM models, with a different value in each coherent patch of size $l_c$ in Eq.~\eqref{eq:coherenced}. And, finally, we perform a global analysis of 20 pulsars measured by the Parkes PTA and of the Crab pulsar measured by the experiment QUIJOTE for $\sim$4.5 years. This allows us to circumvent statistically our lack of knowledge of the exact values of the environmental variables by sampling their known probability distributions. In addition, it allows us to exclude possible harmonic signals of a more prosaic origin, such as source-specific astrophysical phenomena. Including two different experiments in the analysis is also important, to search for effects appearing at different frequencies and for excluding possible signals of instrumental origin.

All in all, our analysis leads to the strongest combined constraints on the ALP ULDM scenario for the range of masses $10^{-23}\lesssim m_{a}\lesssim 2\times 10^{-21}$ eV, as shown in Fig.~\ref{fig:constraints1}. In sec.~\ref{sec:ALPDMbirefringence} we discussed the role of these bounds for models of ALP DM, where the relation between the coupling $g_{a\gamma}$ and the mass $m_a$ (connected to the ALP decay constant $f_a$), is encapsulated by $\kappa$, considered a free parameter.

Finally, by using an idealized model of a pulsar near the Galactic center in a region with a high energy density of DM (and also pulsars in globular clusters), we have highlighted some prospects for probing further the ALP parameter space. \bigskip

The code to reproduce these analyses can be found on \href{https://github.com/jorgetc16/ALP_meter}{GitHub \faGithub}.

\section*{Acknowledgements}

We have used CERNLIB~\cite{cernlib}, NumPy \cite{harris2020array} and Matplotlib \cite{Hunter:2007} for our computations. We thank Benjam\'in Grinstein, Javier Olivares, Mitesh Patel  and Andr\'es Vicente Ar\'evalo for helpful discussions on different aspects of this work. We also thank  Lijing Shao and Wenming Yan for kindly providing us with the data of the time series measured by PPTA.
JMC and Andr\'es Castillo acknowledge support from the Spanish MINECO through the “Flavor in the era of the LHC” grant PGC2018-102016-A-I00 and JMC also from the ``Ram\'on y Cajal'' program RYC-2016-20672. The work of JTC is supported by the Ministerio de Ciencia e Innovaci\'on under FPI contract PRE2019-089992 of the SEV-2015-0548 grant. AC is supported by the Foreign Postdoctoral Fellowship Program of the Israel Academy of Sciences and Humanities. AC also acknowledges support from the Israel Science Foundation (Grant 1302/19), the US-Israeli BSF (Grant 2018236) and the German-Israeli GIF (Grant I-2524-303.7). IFAE is partially funded by the CERCA program of the Generalitat de Catalunya. DB is supported by a `Ayuda Beatriz Galindo Senior' from the Spanish `Ministerio de Universidades', grant BG20/00228.
The research leading to these results has received funding from the Spanish Ministry of Science and Innovation (PID2020-115845GB-I00/AEI/10.13039/501100011033). We thank the staff of the Teide Observatory for invaluable assistance in the commissioning and operation of QUIJOTE.
The {\it QUIJOTE} experiment is being developed by the Instituto de Astrofisica de Canarias (IAC),
the Instituto de Fisica de Cantabria (IFCA), and the Universities of Cantabria, Manchester and Cambridge.
Partial financial support was provided by the Spanish Ministry of Science and Innovation
under the projects AYA2007-68058-C03-01, AYA2007-68058-C03-02, AYA2010-21766-C03-01, AYA2010-21766-C03-02,
AYA2014-60438-P,  ESP2015-70646-C2-1-R, AYA2017-84185-P, ESP2017-83921-C2-1-R, PID2019-110610RB-C21, PID2020-120514GB-I00, IACA13-3E-2336, IACA15-BE-3707, EQC2018-004918-P, the Severo Ochoa Programs SEV-2015-0548 and CEX2019-000920-S, the Maria de Maeztu Program MDM-2017-0765 and by the Consolider-Ingenio project CSD2010-00064 (EPI: Exploring the Physics of Inflation). This project has received funding from the European Union's Horizon 2020 research and innovation program under grant agreement number 687312 (RADIOFOREGROUNDS).

\appendix

\section{Generalized Lomb Scargle Periodogram}
\label{sec:periodogram}

Here we define the Lomb-Scargle periodogram $P_{LS}(\nu)$. Consider a time series with $N$ observations, where $\boldsymbol{y}=\{y_{i}\}$, $\boldsymbol{\sigma}=\{\sigma_{i}\}$ and $\boldsymbol{t}=\{t_{i}\}$ are the data, errors, and time ``vectors'' respectively. The normalized errors weights are
\begin{align*}
w_{i}=\frac{1}{W}\frac{1}{\sigma_{i}^{2}},    
\end{align*}
where $W=\sum1/\sigma_i^2$, and with $\sum w_i=1$. We denote for each frequency $\mathbf{C}=\{c_{i}\}=\{\cos\left(2\pi\nu t_{i}\right)\}$ and $\mathbf{S}=\{s_{i}\}=\{\sin\left(2\pi\nu t_{i}\right)\}$. In addition, we define the products \cite{Zechmeister:2009js,Ivanov:2018byi}
\begin{align*}
Y=\boldsymbol{w} \cdot \boldsymbol{y}, \hspace{0.2 cm} C=\boldsymbol{w} \cdot \mathbf{C}, \hspace{0.2 cm} S=\boldsymbol{w} \cdot \mathbf{S}, \text{ and } 
D&= C_{C} \cdot S_{S} - C_{S} \cdot C_{S}.
\end{align*}
where 
\begin{align*}
Y_{Y}&=\sum_{i=1}^{N} w_{i} y_{i}^2- Y \cdot Y\\
Y_{C}&=\sum_{i=1}^{N} w_{i} y_{i} c_{i}- Y \cdot C, \\
Y_{S}&=\sum_{i=1}^{N} w_{i} y_{i} s_{i}- Y \cdot S,\\
C_{C}&=\sum_{i=1}^{N} w_{i} c_{i}^2- C \cdot C,\\
S_{S}&=1-\sum_{i=1}^{N} w_{i} c_{i}^2- S \cdot S,\\
C_{S}&=\sum_{i=1}^{N} w_{i} c_{i} s_{i}- C \cdot S .
\end{align*}
The generalized LS periodogram with errors weights is then determined as
\begin{align*}
P_{LS}(\nu) = \frac{1}{Y_Y \cdot D} \left( S_S\cdot Y_C \cdot Y_C + C_C \cdot Y_S
\cdot Y_S -2\, C_S \cdot Y_C \cdot Y_S \right),
\end{align*}
where $0\leq P_{LS}(\nu)\leq 1$. Note that the dot product between two ``vectors'' (in bold face) expresses their scalar product and between two scalars it expresses their multiplication.

An alternative way to understand the LS periodogram is as a statistical minimization analysis given by fitting an harmonic model~\cite{Zechmeister:2009js,2018ApJS..236...16V} 
\begin{equation}
\label{eq:LSharmonic}
    y=a_\nu\cos(2\pi\nu t) + b_\nu\sin(2\pi\nu t) + c_\nu, 
\end{equation} 
where the amplitudes $a_\nu$, $b_\nu$ and $c_\nu$ are determined  by a $\chi^2$ minimization process for the different frequencies $\nu$  considered. The inclusion of the term $c_\nu$ generalises the analysis for data with an offset. With this procedure, 
one can recover the generalized LS periodogram, $P_{LS}(\nu)$ as
\begin{equation}
   P_{LS}(\nu)=\frac{1}{2}\left(\hat\chi^2_0-\hat\chi^2_\nu\right),
\end{equation}
where $\hat\chi_\nu$ is the minimum value for that frequency $\nu$ and  $\hat\chi_0$ is the minimum value for a non-relevant reference model.  
 
\section{Effects of the ionospheric corrections}
\label{sec:ionosph}

\begin{figure}[h!]
\begin{subfigure}{.33\textwidth}
  \centering
\includegraphics[width=1.1\textwidth]{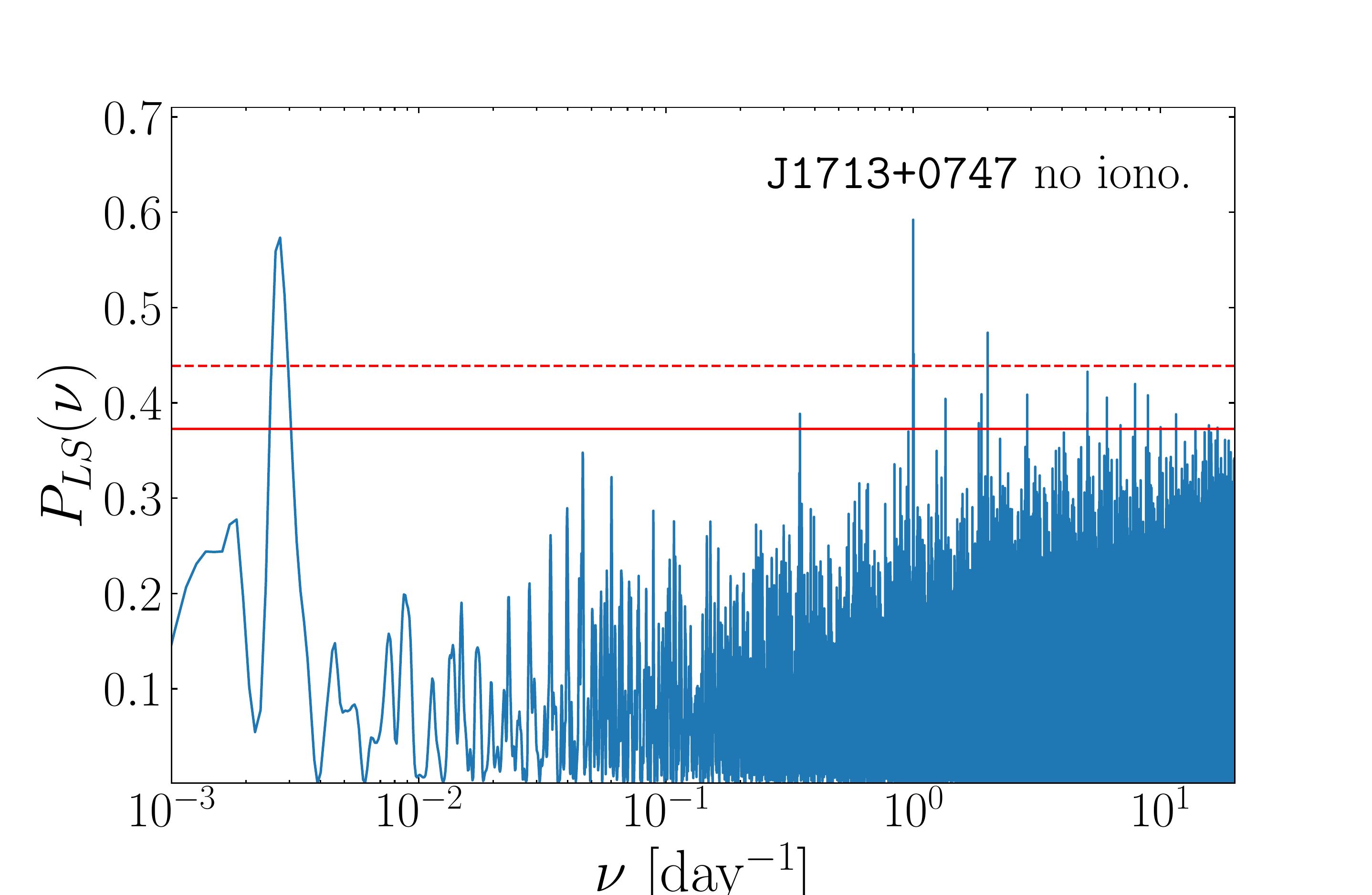}
\end{subfigure}%
\begin{subfigure}{.33\textwidth}
  \centering
\includegraphics[width=1.1\textwidth]{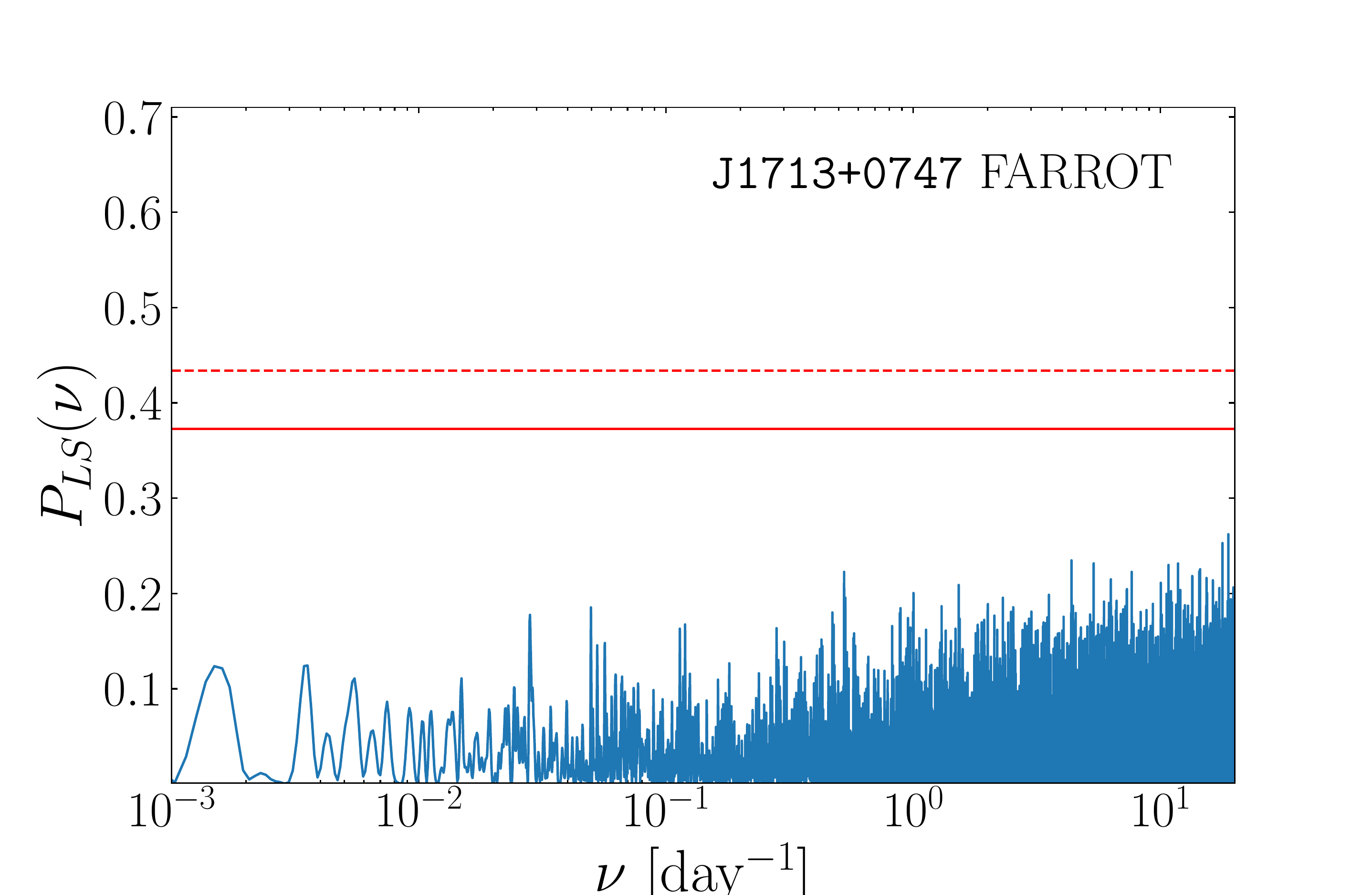}
\end{subfigure}
\begin{subfigure}{.33\textwidth}
  \centering
\includegraphics[width=1.1\textwidth]{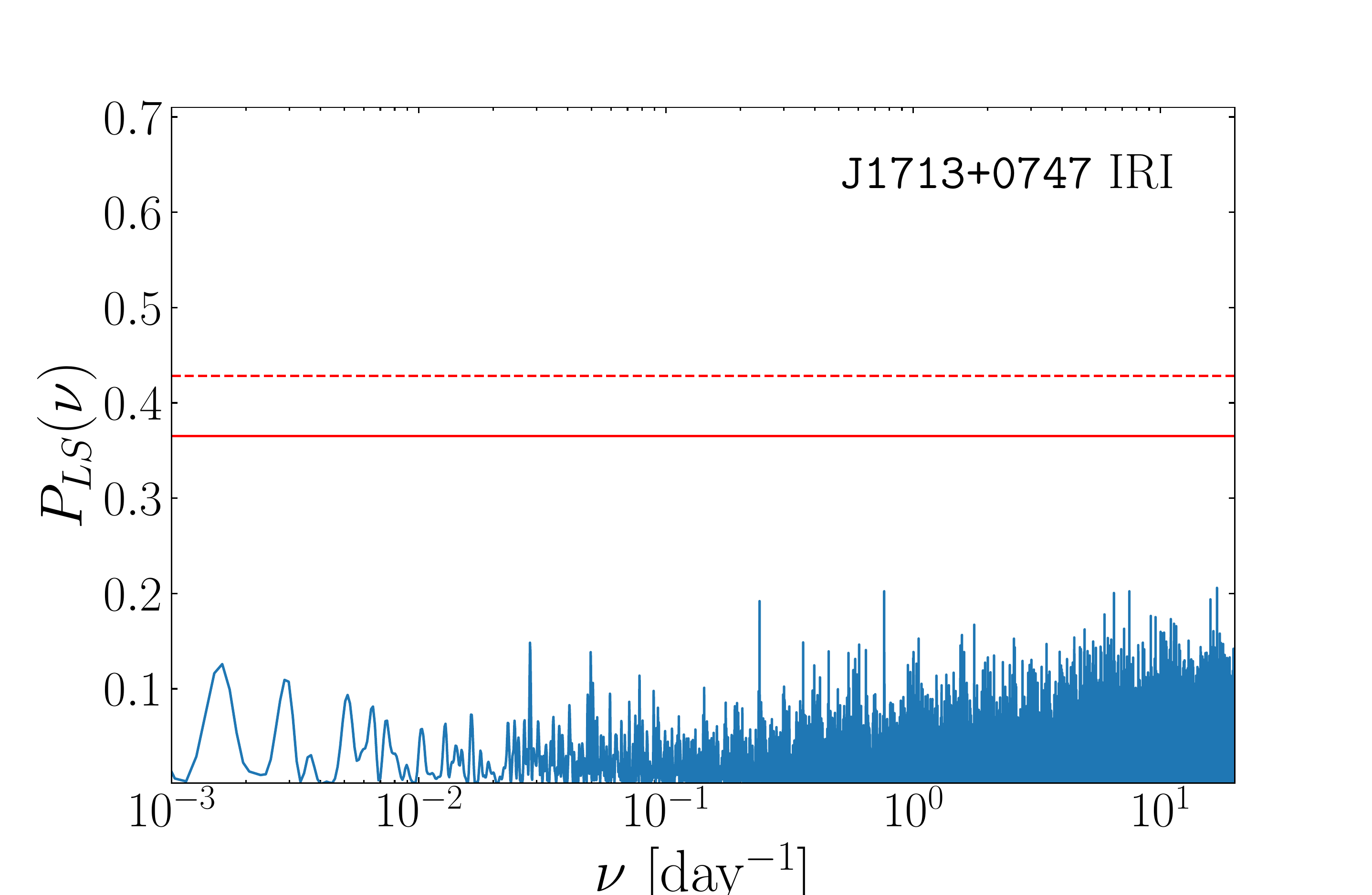}
\end{subfigure}

\begin{subfigure}{.33\textwidth}
  \centering
\includegraphics[width=1.1\textwidth]{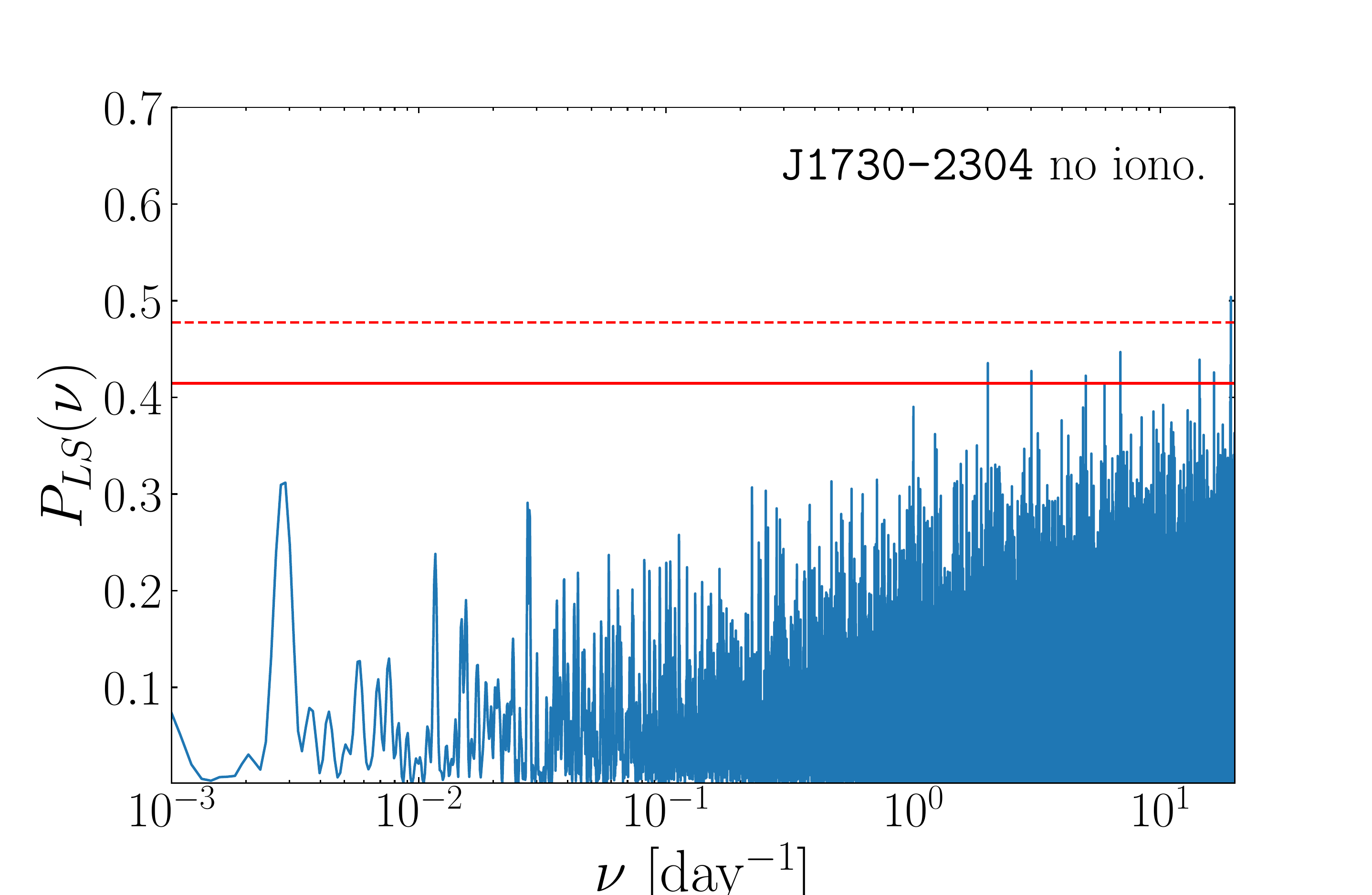}
\end{subfigure}%
\begin{subfigure}{.33\textwidth}
  \centering
\includegraphics[width=1.1\textwidth]{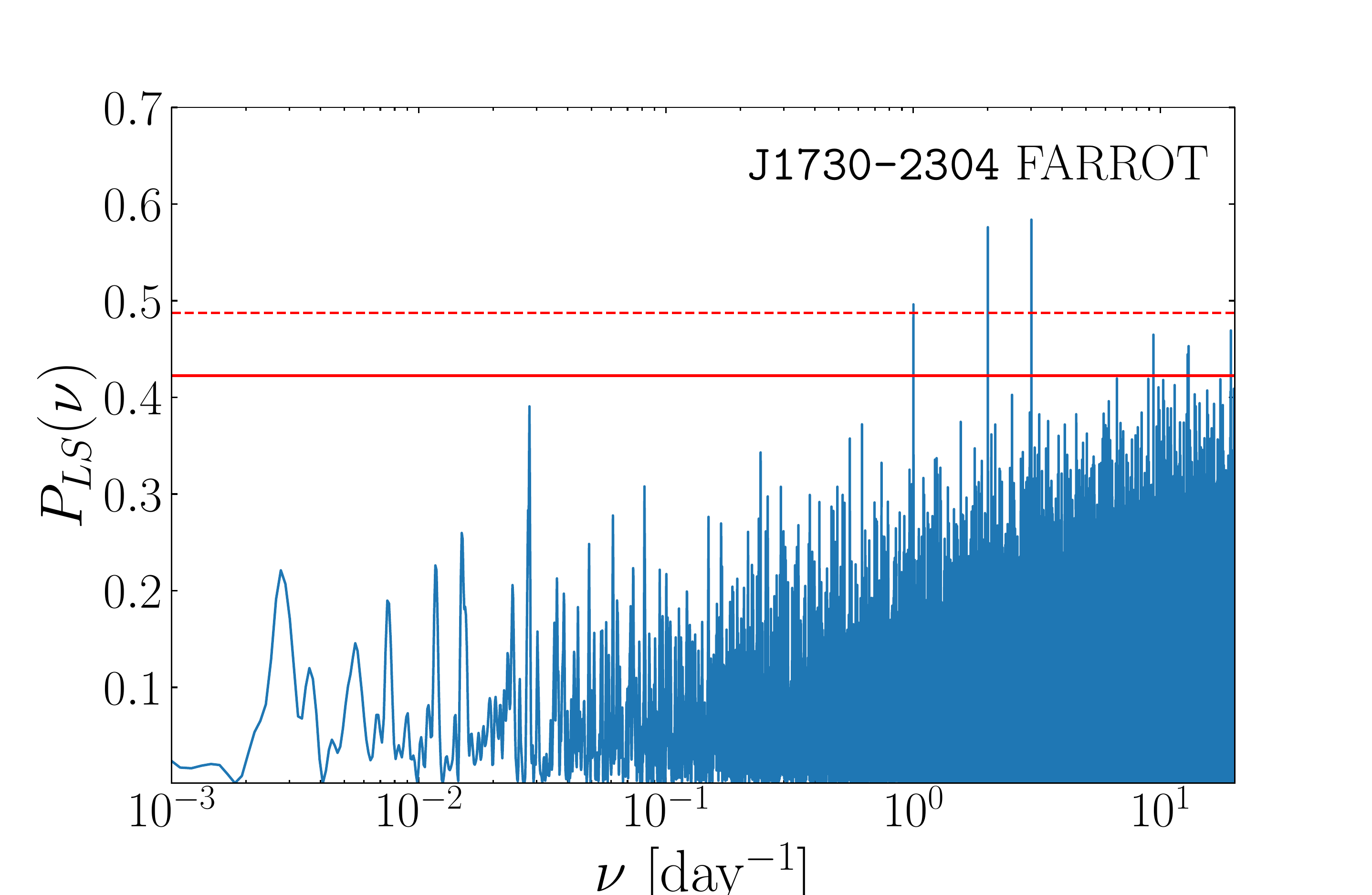}
\end{subfigure}
\begin{subfigure}{.33\textwidth}
  \centering
\includegraphics[width=1.1\textwidth]{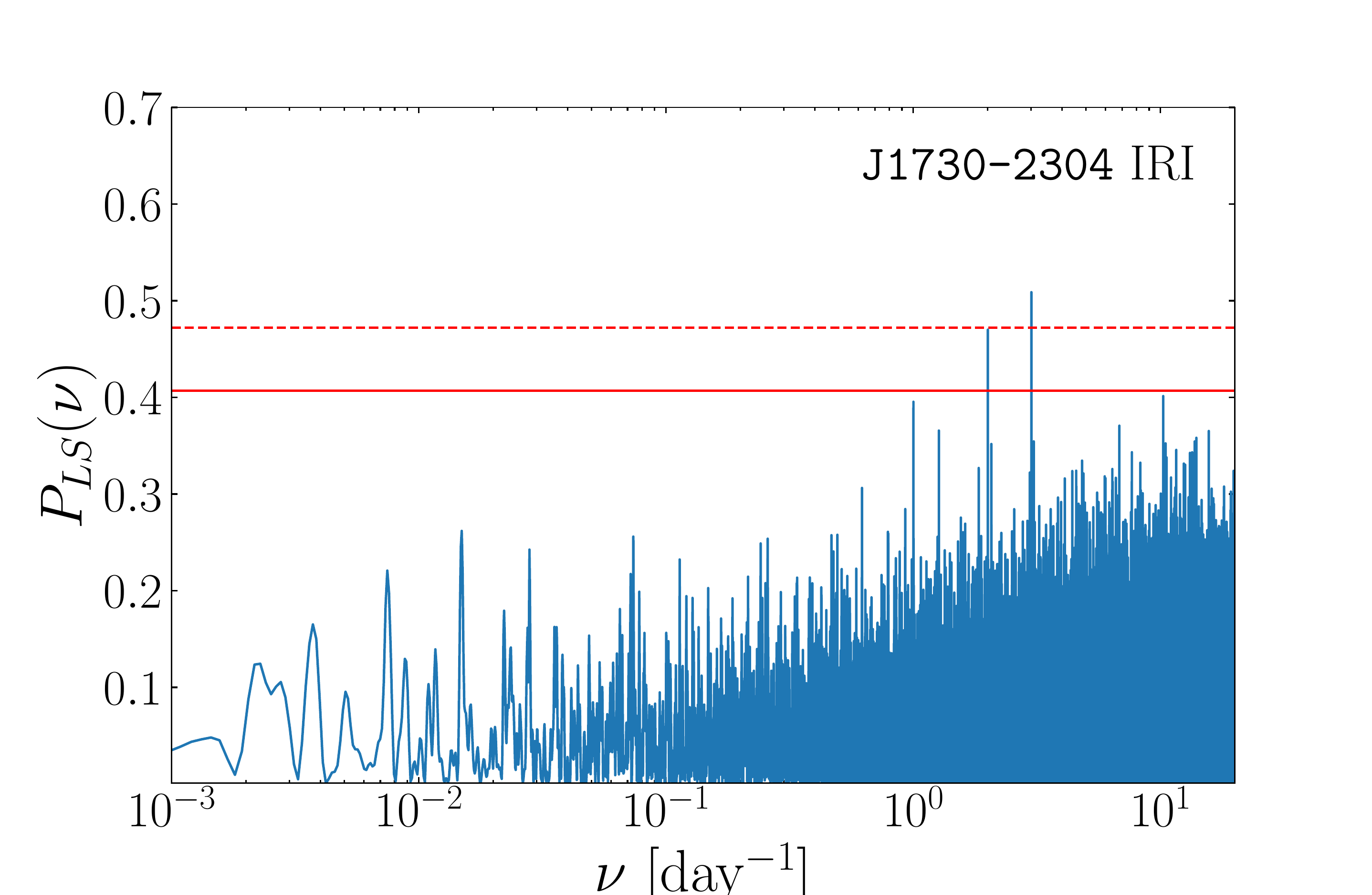}
\end{subfigure}
\caption{Lomb-Scargle periodograms for the time series of polarization measurements for the pulsars \texttt{J1713-0747} (top) and \texttt{J1730-2304} (bottom) without ionospheric corrections (left), with \texttt{FARROT} ionospheric corrections (center) and with IRI ionospheric corrections (right). We also show FAPs at 32\% (red solid line) and 5\% (red dashed line) false-positive rates estimated using a bootstrap method with a 1000 random resamplings of the data set at the same temporal coordinates.}
\label{fig:LSperiodogramsCorrections}
\end{figure}
 
As an example of the impact of \texttt{FARROT} and IRI models of the ionospheric corrections, in Figure \ref{fig:LSperiodogramsCorrections} we show the LS periodograms (see sec.~\ref{sec:methodLS}) for pulsars \texttt{J1713-0747} and \texttt{J1730-2304} without ionospheric corrections and with the two models of corrections discussed above, \texttt{FARROT} and IRI, and the corresponding FAPs at 32\% and 5\%. One can observe that in the case of \texttt{J1713-0747}  without corrections there are two clear peaks over the 5\% FAP line at frequencies corresponding to periods of around one year and one day and some of its aliases that are over the 32\% FAP line, indicating a clear observation of seasonal effects and daily effects. Both peaks clearly disappear once the corrections are applied. Instead in the case of \texttt{J1730-2304}, the modelling of ionospheric corrections seems to introduce ``spurious'' peaks over FAPs of 5\%, which are absent in the original data without ionospheric corrections. Although this affects the background model for the pulsar \texttt{J1730-2304} and its individual bound on $\phi_{95}$, the combined bound exclude these ``fake'' peaks effects in the ALP signals searches. Since the IRI method gives better results in modelling ionospheric corrections (lower seasonal peaks and less spurious ones), we use the data implementing this model in our analyses.

\section{Comparison Stochastic vs. Deterministic analysis}
\label{sec:AppVs}
In general, one might expect a degradation of the bounds on $g_{a\gamma}$ in the stochastic case compared to the deterministic one. 
In order to understand
this effect quantitatively, we perform {\it toy MC} simulations in two limits that are relevant in our analysis: \textit{(i)} {\bf homogeneous}, where the DM density at source and observer are the same $\rho_s=\rho_o$, which is the configuration we approximately have in our dataset (see Tab.~\ref{tab:pulsarlocation}); and \textit{(ii)} {\bf overdense}, when the DM density at the source is much higher than the one at Earth, $\rho_s\gg\rho_o$, which is relevant for our forecasts\footnote{This case also covers the scenario $\rho_s\ll\rho_o$, since the important quantity in~\eqref{eq:polALP2} is the maximum of the energy densities.}. The amplitude of the ALP-induced oscillation in Eq.~\eqref{eq:phi0}, in these two cases and implementing the stochasticity of the ALP field, simplifies as
\begin{align}
\phi^{\rm hom}_a=&\frac{g_{a\gamma}}{\sqrt{2}m_a} \rho_{\rm DM}^{1/2} \left( \alpha_o^2+ \alpha_s^2-2 \alpha_o\alpha_s\cos\Delta\right)^{1/2}, \\
\phi^{\rm over}_a=&\frac{g_{a\gamma}}{\sqrt{2}m_a} \rho_{\rm DM}^{1/2} \alpha_s.
\end{align}
In the deterministic case ($\alpha_i=1$) this simplifies further to
\begin{align}
\phi^{\rm hom}_a=&\frac{g_{a\gamma}}{m_a} \rho_{\rm DM}^{1/2} \left(1- \cos\Delta\right)^{1/2}, \\
\phi^{\rm over}_a=&\frac{g_{a\gamma}}{\sqrt{2}m_a} \rho_{\rm DM}^{1/2}.
\end{align}

We perform several pseudo-experiments where we sample a measurement of $\phi_a$, $\Delta$ and of the $\alpha_i$ and we derive the limits in the deterministic and stochastic cases to compare them. For this toy MC we assume observations $\phi_i$ over $i=1,\ldots,N_s$ sources all featuring the same precision (that is, the same scale factor for the Rayleigh distribution) and with the same $\rho_{\rm DM}$.

Surprisingly, we find that the upper limit on $g_{a\gamma}$ improves in the stochastic analysis by a factor 2.6 in the homogeneous case, while it worsens (as expected) by a factor 0.55 for the overdense one. The improvement in the homogeneous limit seems to be due to a nontrivial interplay between the random $\Delta$ and $\alpha_i$ variables. The impact of stochasticity is intuitively more severe for the overdense case, as the sensitivity of a given pulsar to $g_{a\gamma}$ hinges only on the particular value of $\alpha_s$ (compared to the former where both $\alpha_s$ and $\alpha_o$ are relevant). If we neglect the effect of $\Delta$ by e.g. taking the root mean square value of Eq.~\eqref{eq:phi0} (as often done in the literature~\cite{PDPAxion,Liu:2019brz}) then we would obtain a limit that is worse by a factor 0.92 also in the homogeneous case.

Expanding on the toy example, we repeat the pseudo-experiments for $N_s$ sources, obtaining a ``global'' value of $g_{a\gamma}$ as the weighted average of those measured at each source. 
In the stochastic case,
$g_{a\gamma}\propto (\alpha_o^2+\alpha_s^2-2\alpha_o\alpha_s\cos\Delta)^{-1/2}$, and a larger $\alpha_s$ or $\alpha_o$ will translate into a smaller $g_{a\gamma}$. When this happens, the variance of $g_{a\gamma}$ that determines the weight in the average 
also shrinks, 
giving a higher weight to that measurement. Moreover, in the case of $\Delta\to\pi$, the interference with the $\cos\Delta$ term will be constructive and its effect become greater.  It is not likely to get a high value of $\alpha_s$ and a $\Delta$ close to $\pi$ for a single pulsar; however, as we add more sources the probability of getting at least one with a high value of $\alpha_s$ increases. In the deterministic case,  $g_{a\gamma}\propto (2-2\cos\Delta)^{-1/2}$ is bounded by $2$, but as $N_s$ increases these larger denominators (and therefore smaller $g_{a\gamma}$ and variances) will dominate the weighted average, making the resulting PDF thinner and closer to the stochastic one.

In the overdense scenario, for the stochastic case, 
we have $g_{a\gamma}\propto \alpha_s^{-1}$ (the interference term is subleading). In the deterministic case, the environmental variables have no effect and the PDF will be constant no matter how many pulsars one adds. This is why we see a stronger bound in the deterministic case for a small number of $N_s$ while, as $N_{s}$ increases, the probability of getting a high value of $\alpha_s$ increases, and the stochastic bound will eventually overcome the deterministic one.

All these effects are seen in Table \ref{tab:toyratio}, where we show the size of the bounds at 95\% C.L. obtained in the deterministic scenario relative to those obtained in the stochastic one as a function of $N_s$ in the two cases discussed above. Interestingly, in the homogeneous case the difference becomes small as more sources are added to the analysis. For the overdense case, which is relevant in the forecasts presented in Sec.~\ref{sec:forecasts}, a boost in the sensitivity could be achieved by studying many sources with similar statistical power.

\begin{table}[t]
\centering
\setlength{\tabcolsep}{1.07em}
\begin{tabular}{lcccccccc}
\hline\hline
$N_s$ &1&2&3&4&5&10&25&100\\
\hline
{\bf Homogeneous}&2.60&1.30&1.16&1.13&1.12&1.12&1.12 & 1.11\\
{\bf Overdense}&0.56&0.91&1.05&1.14&1.19&1.33&1.44&1.52\\
\hline\hline
\end{tabular}
\caption{Ratio of the bounds at 95\% C.L. obtained in the deterministic scenario over the ones obtained in the stochastic scenario for different number of sources, $N_s$. \label{tab:toyratio}}
\end{table}
%
\section{Comparison between periodograms and the frequentist method}
\label{sec:comparetofreq}
The frequentist method used in~\cite{Caputo:2019tms} relies on the comparison between the $\chi^2$ obtained for a fit to a function with some parameters, in this case to an harmonic oscillating function, and the $\chi^2$ of the null hypothesis, i.e., assuming any fluctuation in the data is due to statistical noise. A value of the parameters of the fit will be ruled out at certain C.L. when $\Delta\chi^2\equiv\chi^2_{\rm{null}}- \chi^2_{\rm{fit}}>N$, with $N$ depending on the number of parameters of the fit. 

As discussed in Appendix~\ref{sec:periodogram}, the value of the periodogram at a given frequency is equivalent to the result of a minimization of a $\chi^2$ with also a harmonic function at that frequency. However, in the periodogram approach the C.L. regions are derived by using the FAP procedure, and not assuming a $\chi^2$ distribution given by the statistical nominal errors of the data. This is certainly not the case as the $\chi^2/{\rm d.\, o.\, f}$  found for both the null hypothesis and the periodic function are extremely large (e.g. $\chi^{2}/\rm d.\, o.\, f$ is $488.6$ for the polarization angle of \texttt{J0437-4715} measurements using IRI corrections), indicating that none of them is a good fit to the data. 
Namely, for some combination of the parameters, one finds that the $\chi^2_{\rm{fit}}$ is significantly smaller than $\chi^2_{\rm{null}}$, from which one could deduce that the presence of a periodic signal is favoured by the data with a very high frequentist C.L. Still, both fits are disfavoured by the large $\chi^2/{\rm d.\, o.\, f}$, which should   be taken into account in the final bounds. On the other hand, the periodogram approach, with FAPs estimated using bootstrap methods, provides a robust estimate of the C.L. for the presence or absence of harmonic signals in the time series of the polarization measurements.     

\begin{figure}[ht]
\begin{center}
\begin{tabular}{cc}
\includegraphics[width=0.75\textwidth]{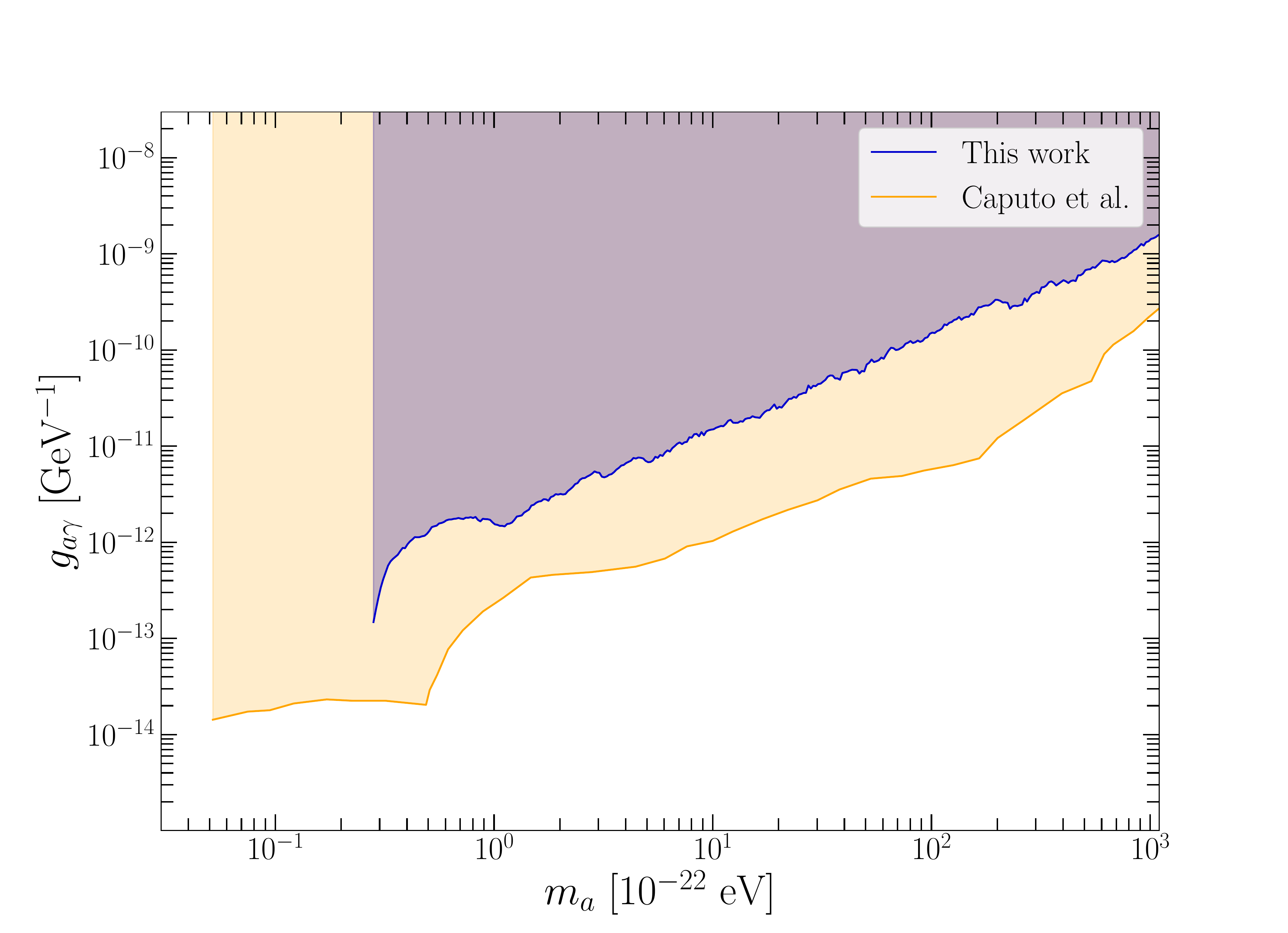} 
\end{tabular}
\caption{Bounds on the photon-ALP coupling at 95$\%$ C.L. derived with the pulsar \texttt{J0437-4715}. The orange area corresponds to the bounds derived with the method in \cite{Caputo:2019tms}, while the blue region is derived in this work (under the deterministic assumption). \label{fig:comparison}}

\end{center}
\end{figure}

In Fig.~\ref{fig:comparison} we compare the deterministic limits reported for the \texttt{J0437-4715} using the frequentist approach in~\cite{Caputo:2019tms} with the ones that we obtain using the periodograms. One can see that the frequentist interpretation of the C.L. can lead to limits an order of magnitude stronger than the more realistic ones given by the FAPs. 

In terms of the range of frequencies considered, for \texttt{J0437-4715} the range of masses studied in \cite{Caputo:2019tms} goes down to $m_a=5\times 10^{-24}$ eV, which is equivalent to a frequency of $\nu=10^{-4}\, {\rm days^{-1}}$ or a period of $T=26.6$ years. As discussed in Sec.~\ref{sec:methodLS}, periodic harmonic signals with such low frequencies can be confused with a linear signal. Since our method critically relies on the study of periodic signals in the data, it is not possible for us to  explore the ultra low mass regime with periodograms, as we can see in Fig.~\ref{fig:comparison}.

Finally, in \cite{Caputo:2019tms}, only one of the available pulsars was taken into account. As  discussed in the main text, using several sources is important,  not only because the signal of ULDM at a given frequency should appear in all sources, but because it diminishes the stochastic effects that affect the constraint.

\bibliographystyle{unsrt}
\bibliography{polar}
\end{document}